\documentclass[sensors,article,accept,moreauthors,pdftex]{Definitions/mdpi} 

\usepackage{gensymb}
\usepackage{siunitx}
\usepackage[none]{hyphenat}
\usepackage{soul}
\firstpage{1} 
\pubvolume{1}
\issuenum{1}
\articlenumber{0}
\pubyear{2022}
\copyrightyear{2022}
\externaleditor{Academic Editors: {Rosa Ma Alsina-Pagès, Giovanni Zambon and Antonio Lázaro}} 
\datereceived{29 November 2021} 
\dateaccepted{6 February 2022} 
\datepublished{} 
\hreflink{https://doi.org/} 

\newif\ifrev
\revtrue 
\ifrev 
\else

\fi

\Title{Evaluation of Acoustic Noise Level and Impulsiveness Inside Vehicles in Different Traffic Conditions}

\TitleCitation{Evaluation of Acoustic Noise Level and Impulsiveness Inside Vehicles in Different Traffic Conditions}

\Author{Daniel Flor $^{1,}$*\orcidB{}, Danilo Pena $^{2}$\orcidA{}, Hyago Lucas Oliveira $^{1}$\orcidF{}, Luan Pena $^{2}$\orcidC{}, Vicente A. de Sousa, {Jr.} $^{1}$\orcidE{} and~Allan~Martins~$^{2}$\orcidD{}} 

\AuthorNames{Daniel Flor, Danilo Pena, Hyago Lucas Oliveira, Luan Pena, Vicente Angelo, Allan Martins}

\AuthorCitation{Flor, D.; Pena, D.; Oliveira, H.L.; Pena, L.;  de Sousa, V.A., Jr.; Martins, A.}

\address{%
$^{1}$ \quad Department of Communications Engineering, Federal University of Rio Grande do Norte, Natal~59078-970,~Brazil; hyago.oliveira.119@ufrn.edu.br (H.L.O.); vicente.sousa@ufrn.br (V.A.d.S.J.)\\
$^{2}$ \quad Department of Electrical Engineering, Federal University of Rio Grande do Norte, Natal 59078-970, Brazil; danilo@dca.ufrn.br (D.P.); luan.gppcom@gmail.com (L.P.); allan@dca.ufrn.br (A.M.) }

\corres{Correspondence: danielflor@ufrn.edu.br}

\abstract{Recently, the issue of sound quality inside vehicles has attracted interest from both researchers and industry alike due to health concerns and also to increase the appeal of vehicles to consumers. This work extends the analysis of interior acoustic noise inside a vehicle under several conditions by comparing measured power levels and two different models for acoustic noise, namely the Gaussian and the alpha-stable distributions. Noise samples were collected in a scenario with real traffic patterns using a measurement setup composed of a Raspberry Pi Board and a microphone strategically positioned. The analysis of the acquired data shows that the observed noise levels are higher when traffic conditions are good. Additionally, the interior noise presented considerable impulsiveness, which tends to be more severe when traffic is slower. Finally, our results suggest that noise sources related to the vehicle itself and its movement are the most relevant ones in the composition of the interior acoustic noise.}


\begin{document}
\section{Introduction}

Acoustic noise has received much attention in the automotive industry due to the increasing demand for in-vehicle voice assistant systems~\cite{wp_voicebot}. Noise evaluation is an essential issue in this field, enabling the design of in-vehicle multimedia systems with better noise control and fewer disturbances that degrade acoustic communications performance in a vehicle interior. Such disturbances may have different sources, and identifying them in order to focus on the most dominant sources will result in more efficient noise controls and optimized systems for audio applications. The statistical characteristics of the noise are crucial to define and configure the active noise control techniques~\cite{Chen2021}.

Therefore, this work presents an experimental evaluation of the characteristics of the acoustic noise inside a vehicle under the perspective of an in-vehicle voice reception system. We also capture sources related to the traffic that might impact a car's interior environment, providing insights concerning the acoustic noise and its source in the vehicle interior.



{Our previous analysis~\cite{Flor_2020} showed how some factors such as traffic are correlated to the average noise power inside the vehicle. For that evaluation, a total of 194 noise samples were collected using a measurement setup installed in a C4 Lounge from Citroën. In this paper, we extended our previous work on the subject in the following ways:}

\begin{itemize}
    \item {Acquiring an additional 254 noise samples, which were collected in a different vehicle and a different time of the year from our previous work~\cite{Flor_2020};}
    \item {Validating our setup using a sound pressure level meter to verify the power levels measured;}
    \item {Improving the statistical evaluation of the selected variables, examining which of them have more influence in the noise power levels inside, comparing the results between both measurement sets, and providing a more in-depth analysis of the effect of the car's windows;}
    \item {Evaluating the degree of impulsiveness of the interior noise and comparing the AWGN and alpha-stable noise models;}
    \item {Analyzing the window size for the estimation of alpha-stable distribution parameters.}

\end{itemize}

{Finally, we highlight that the complete collection of measurements, including information about the conditions and location, is freely available~\cite{noise_data2019} and can be helpful for different purposes.}

This paper is organized as follows. Section~\ref{sec:levantamento} presents a brief overview of the literature on vehicle interior noise. The measurement campaigns and setup are described in Section~\ref{sec:measurement}. Section~\ref{sec:methods} presents the statistical methods used to analyze the collected data, and Section~\ref{sec:results} discusses the results. Finally, in Section~\ref{sec:conclusions}, we present our final remarks.

\section{Related Works}
\label{sec:levantamento}

The topic of acoustic noise or sound quality in vehicles is a multidisciplinary subject that is related not only to the health and comfort of drivers and passengers but also to the appeal of vehicles as a product. Thus, since the beginning of the car industry, many studies have been developed exploring different aspects of sound quality. Many of these studies focus on the effects of traffic noise on human health. There is research on the impact of noise on sleep and mental health~\cite{BASNER2014,Kara2019,SYGNA2014}, on the development of cognitive processes in children~\cite{Gary2003}, and on the increase of risk of heart diseases~\cite{ROSWALL2017,VIENNEAU2015} and diabetes~\cite{ROSWALL2018}. Recent studies present contributions to identifying and optimizing vehicle interior noise~\cite{Huang2020_2,Liu2021,Horne2021,Ningning2020,Lu2021,Jianghua2018}, approaching different noise sources. The sound quality of the vehicle cabin is also essential for various in-vehicle applications, such as multimedia \cite{Almalkawi2010,Wang2007,Jennehag2016}, security~\cite{Toma2019,Kordov2019,Liu2019}, assistive~\cite{Kuhn99,Higuchi2017,Khan2017}, and autonomous vehicles ~\cite{Tremoulet2019,Chen2019}. Furthermore, several studies in psychoacoustics seek to establish objective metrics to assess the subjective sound quality perceived by vehicle passengers~\cite{DOLESCHAL2021,LIAO2020}, which is an essential factor for consumer satisfaction and the marketability of a vehicle~\cite{SWART2019}.

Given the importance of the subject, several studies were developed to characterize internal noise in vehicles. As the noise perceived inside the cabin is a composition of noise sources of different natures, such as wind, engine, and rolling, most works focus on describing the contribution of specific components. Knowledge of the most relevant sources can indicate the main challenges in acoustic systems and the best way to represent them mathematically. Therefore, its characterization is essential for anyone interested in vehicular acoustic systems.

According to the literature~\cite{Huang2018,Jung2019}, the different contributions to in-vehicle noise can be classified according to their source. Noise can originate from the structural vibrations of the car and its components or from aerodynamic excitations~\cite{vanherpe2011} transmitted by the cabin of the car. For instance, the noise created by the tires/road interaction is usually separated into two components~\cite{Jung2019}: structural low-frequency noise (below 500 Hz)~\cite{PRATICO2014,DELPIZZO2020}, and aerial noise, with medium and high-frequency contributions (above 500 Hz). Understanding the characteristics of this particular noise source is essential for the development of low-noise roads~\cite{LICITRA2017,PRATICO2012,LICITRA2015}. Other works investigate the sound quality of specific vehicle phenomena and components, such as closing car doors~\cite{Shin2013,Parizet2008}, engine noise~\cite{Li2018,Huang2018}, Heating, Ventilation, Air Conditioning (HVAC) systems~\cite{Soeta2017}, seat belts~\cite{christopher2004}, and wind~\cite{YOSHIDA2017,Talay2019}.

Some studies use linear regression modeling~\cite{Volandri2018} to establish correlations between objective psychoacoustics metrics, such as pitch, roughness, volume, and others~\cite{Wang2014}, and subjective sound quality metrics, which are often obtained from jury reviews~\cite{Wang2014}. With a similar aim, some works use machine learning techniques, such as clustering and neural networks, to model the contribution of one or multiple sources ~\cite{Pietila2012,Huang2020,Park_Kang_2020} or even the human auditory system~\cite{Wang2017}. However, these models usually use psychoacoustics metrics, focusing on predicting the assessment of the subjective perception of the sound quality of a passenger or driver. Many works contribute to noise prediction using artificial intelligence, providing a model for the traffic noise~\cite{PuyanaRomero2016,Bravo-Moncayo2019,Yin2020} and sound quality prediction~\cite{Huang2021,Qiu2021,Wang2021} contexts.

Although there are a wide variety of works on the evaluation of interior vehicle noise, most of them focus on studying one or a few noise sources at a time~\cite{Huang2018}. The works on modeling and prediction are usually elaborated from the perspective of psychoacoustics, whose metrics may not be relevant for a voice processing system in a vehicular multimedia center. In addition, these studies are often conducted in laboratory or highly controlled environments~\cite{SWART2019,Huang2020,Liu_2020}. Thus, relevant sources of acoustic noise present in a real driving scenario in an urban environment, and the composition of their effects are potentially disregarded.


Additionally, in our literature survey, studies that seek to consider the composition of multiple sources for in-vehicle noise with probability distribution modeling were not found. One of the most popular noise models in communication systems is the Additive White Gaussian Noise (AWGN) model~\cite{Kristoffer2018}, based on the Gaussian probability distribution. Although the AWGN model is of great importance, it is not always adequate~\cite{Nikias_1994}. Impulsive phenomena, in which the noise changes suddenly to a value far from the mean in a short period of time, can affect the performance of signal processing solutions based on traditional Gaussian modeling~\cite{Yardimici_1998,Georgiou1999,Georgiou_2006}. In the context of this work, impulsive noise is relevant to source location~\cite{Georgiou_2006,Guo_2021,Mario2020,Mario2020_2,Danilo2020}, voice processing~\cite{Liu_2017,Naoto_2020}, and noise comfort and pollution~\cite{RAJALA_2020,Schaffeld_2020}. In several of the works cited above, impulsiveness is modeled by an alpha-stable distribution. Alpha-stable distributions are widely used to represent a range of phenomena for which non-Gaussian behavior is expected. The flexibility of its parameters, which allow for changes in the symmetry, dispersion, and tail mass of the distribution, as well as the Generalized Central Limit Theorem and empirical evidence~\cite{Nolan2020} justify the use of stable models in applications such as econometrics, computing, meteorology, medicine, and image processing, among others~\cite{Nolan_bib,Nolan2020}.

\section{Measurement Campaigns and Setup}
\label{sec:measurement}

\subsection{Measurement Campaigns}
\label{sec:campaign_conditions}

The evaluated interior noise data were obtained in two separate measurement campaigns. Both campaigns took place in Natal, Brazil. Located in northeastern Brazil, the city has an area of \SI{167}{km^2} and a typical tropical climate with warm temperatures and high humidity throughout the year. The first campaign was carried out in June and July of 2019. The samples collected and the result of their analysis were presented in our previous work~\cite{Flor_2020}. To extend the results of the previous work, we also conducted a second measurement campaign, which occurred between April and May of 2021. It is worth emphasizing 
that the second campaign was carried out during the SARS-CoV-2 pandemic. Due to social distancing measures such as the closing of schools, restaurants, and public spaces, the traffic patterns in the city were altered. In particular, during the weeks with high-level restrictions, traffic was less heavy than expected in some regions of the city.

Even though both campaigns use the same measurement setup and have the same objectives, differences between the results for each are expected. Factors such as the model of the cars, the driver, the months of the year, 
and different traffic patterns, among others, can influence the interior noise and the data collection process. In this work, we aim to evaluate if the results for both campaigns are compatible and exhibit the same trend, especially considering the higher amount of measurement points in the second campaign.

\begin{figure}[H] 
\includegraphics[width=0.9\linewidth]{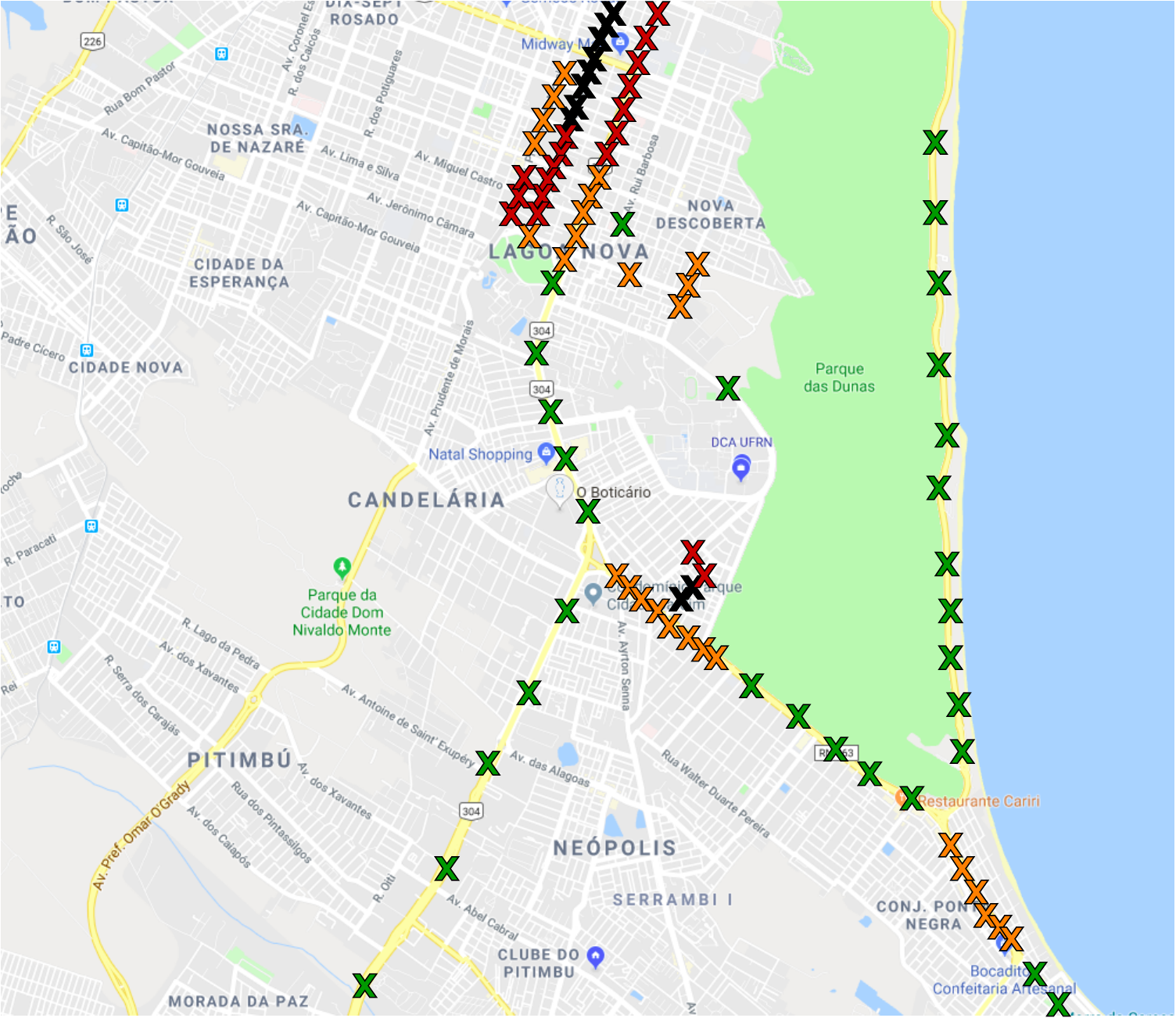}
\caption{Map of Natal 
with markings for each measurement point in the first measurement campaign. The colors indicate the traffic condition at the time of the measurement, {as described in Table~\ref{Tab:variaveis_trafego}}.}
\label{fig:map1}
\end{figure}

\begin{figure}[H] 
\includegraphics[width=0.78\linewidth]{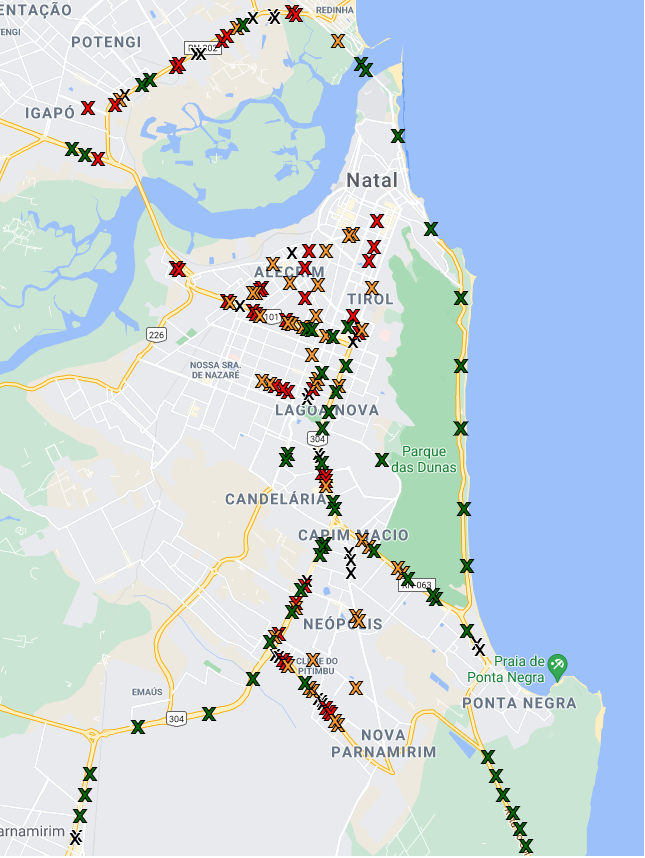}
\caption{Map of Natal 
with markings for each measurement point in the second measurement campaign. The colors indicate the traffic condition at the time of the measurement, {as described in~Table~\ref{Tab:variaveis_trafego}}.}
\label{fig:map2}
\end{figure}

The sampling points were located in different streets and avenues, aiming to spread uncontrolled conditions such as crowd noise. Figures~\ref{fig:map1} and~\ref{fig:map2} show the sampling locations. The colors in the marking represent the traffic conditions associated with each sample. These conditions are defined following Google Maps'  traffic conditions policy~\cite{Pokorny2017}. There are four possible traffic conditions, according to the mean speed of the cars in the street. Table~\ref{Tab:variaveis_trafego} lists the traffic categories and the speed intervals used by the map application to classify the traffic in each street. Moreover, the car used in the data acquisition was always at the speed interval for that specific measurement. The exact car speed value can be found in our dataset~\cite{noise_data2019}. It is also expected that on average, the nearby cars are in the same speed interval.

\vspace{-6pt}

\begin{table}[H] 
\caption{Description of the\textit{Traffic} variable.}  

\label{Tab:variaveis_trafego}
\resizebox{1\linewidth}{!}{%
\begin{tabular}{@{}ccc@{}}
\toprule
\textbf{Traffic Condition (Color)} & \textbf{Speed Interval}         & \textbf{Description}                                          \\ \midrule
Black                              & 0                                        & Indicates extremely slow traffic.                       \\
Red                           & \textless 20 km/h                        & Traffic moves slowly. \\
Orange                            & \textgreater 20 km/h and \textless 40 km/h &  Intermediate traffic flow.                                  \\
Green                              & \textgreater 40 km/h                      & Indicates that traffic is fast.                              \\ \bottomrule
\end{tabular}%
}
\end{table}

In addition to the traffic condition, each of the sampling points has different characteristics. Hence, we measured the noise in many different areas of the city to represent the different noise sources of each environment. We aimed to measure each traffic category in different streets and times of the day. For example, for the \textit{Green} condition, we collected samples in the federal highway BR-101, which always presents fast but intense traffic flow with multiple lanes and also measured in the coastal highway, which comparatively has fewer cars and lanes but presents sources such as the wind and the ocean. In addition, the measurements were done at different times of the day for each location in order to represent the variations in traffic patterns throughout the day.  

All measurements were obtained in asphalt with smooth road surface conditions with no potholes or unevenness. The HVAC systems were turned off. The participants were quiet during measurement, and all objects that could create noise during the car's movement were removed. Furthermore, to avoid bias in the impulsiveness analysis, we removed from the dataset some of the samples that could represent outliers. We checked each measured audio sample for highly impulsive events that are not part of the observed variables or that could not be adequately represented by the amount of collected samples:


\begin{itemize}
    \item Car horns;
    \item Potholes or unevenness in the road;
    \item Speed bumps;
    \item Sudden braking or acceleration due to traffic, intersections, or traffic lights;
    \item Excessive noise from heavy vehicles;
    \item Noise from multiple motorcycles passing close by;
    \item Music and advertisements from other cars or establishments on the street;
    \item People talking around the vehicle;
    \item Noise from animals such as dogs, birds, and cicadas;
    \item Ambulance and police car sirens;
    \item Unidentified noise sources and other events.
\end{itemize}

While one can argue that many of the events listed above are common in a typical traffic scenario, it should be noted that the main objective of this work is to analyze the contribution of the controlled variables to the noise level inside a car and how these variables affect the impulsiveness of this noise. For example, a car horn is an event that will be impulsive and contribute significantly to the noise observed inside the car regardless of the traffic conditions.

\subsection{Controlled and Uncontrolled Variables}
\label{sec:controlled_variables}
In addition to the traffic condition explained above, two other variables were controlled during the campaigns, as presented in Table~\ref{Tab:variaveis_ambiente}. 
They are the position of the car windows and the maximum speed of the car at the moment during a measurement. These variables, along with traffic conditions, were chosen based on our literature review and due to being easily controllable. Care was taken to obtain noise samples for all combinations of these variables. 
The car speed was always compatible with the speed interval of the traffic conditions described in Table~\ref{Tab:variaveis_trafego}.

\begin{table}[H] 
\caption{Description of the controlled environment variables.}
\label{Tab:variaveis_ambiente}
\resizebox{\textwidth}{!}{%
\begin{tabular}{@{}ccc@{}}
\toprule
\textbf{Variable} & \textbf{Possible Values} & \textbf{Notes}                 \\ \midrule
Windows positions & Open; Closed              & All four windows on the same position. \\
Traffic & Black; Red; Orange; Green & Speed interval (see Table~\ref{Tab:variaveis_trafego}). \\
Speed        & 0--80 km/h                   & Maximum value during measurement interval.                                 ~~ \\ \bottomrule
\end{tabular}
}
\end{table}

The three variables in Table~\ref{Tab:variaveis_ambiente}, along with time and location, are the variables we could control during our experiment. However, each measurement is affected by far more variables. The model of the car, the type of road, the weather, the driver, the number of people on the streets, and many other factors can affect the noise characteristic inside the vehicle. Some of these variables have fixed values (such as the car model or the absence of rain). For the other uncontrolled variables, care was taken to represent their effects in the sample data. For instance, we drove through many different streets and avenues to account for variations in the type of asphalt between roads. We expect that the selected controlled variables will significantly affect 
the interior noise levels and \mbox{impulsiveness~\cite{vanherpe2011,Huang2018}.} However, it is worth bearing in mind that other factors not accounted for in this experiment may also influence the noise.

\subsection{Measurement Setup}\label{sec:measurement_setup}

\begin{figure}[H] 
\includegraphics[width=0.62\linewidth]{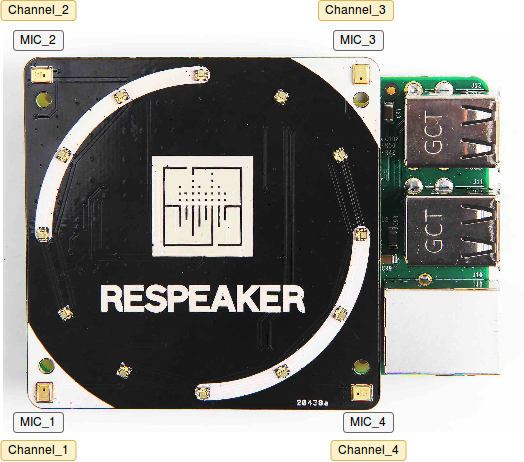}
\caption{RespeakerCore v1 (MT7688) board used to record the acoustic noise inside the car. }
\label{fig:respeaker}
\end{figure}

The measurement setup used is an adaption of the one presented in~\cite{Danilo2019}, which was composed of an Analog-to-Digital Converter (ADC) AC108 embedded in an expansion board for Raspberry Pi, called ReSpeaker Core v1 (MT7688)  board~(Figure~\ref{fig:respeaker}~\cite{respeaker_Online}). The instrument's specifications are described in Table~\ref{Tab:respeaker_specs}. The recordings were stored using a Raspberry Pi 3 (Model B), which also controlled the setup. Each microphone records a five seconds-long audio sample, although only the samples from the first channel were used.


\begin{table}[H] 
\caption{Description of the ReSpeaker Core v1 specifications.}
\label{Tab:respeaker_specs}
\setlength{\tabcolsep}{22.5mm}{\begin{tabular}{cc}
\toprule
\textbf{Specification} & \textbf{Value} \\ \midrule
Microphone channels & 4 \\
ADC model & AC108 \\
Digital output & I2S/TDM \\
System clock & 19.2 MHz \\
Sampling rate & 48 kHz \\ \bottomrule
\end{tabular}
}
\end{table}

\begin{figure}[H] 
\includegraphics[width=0.70\linewidth]{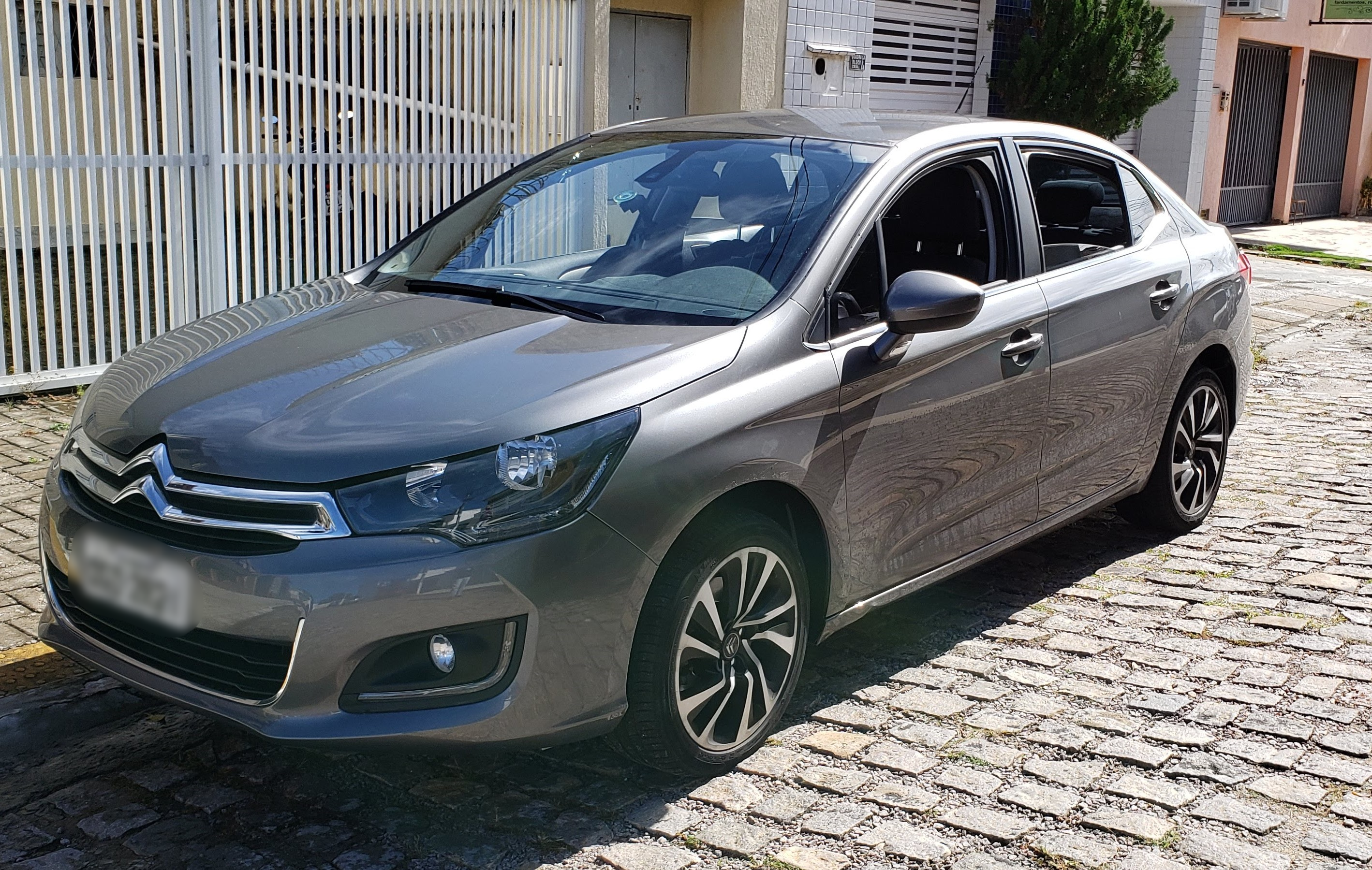}
\caption{Vehicle used in the first campaign.}
\label{fig:carro_1_externo}
\end{figure}

\begin{figure}[H] 
\includegraphics[width=0.70\linewidth]{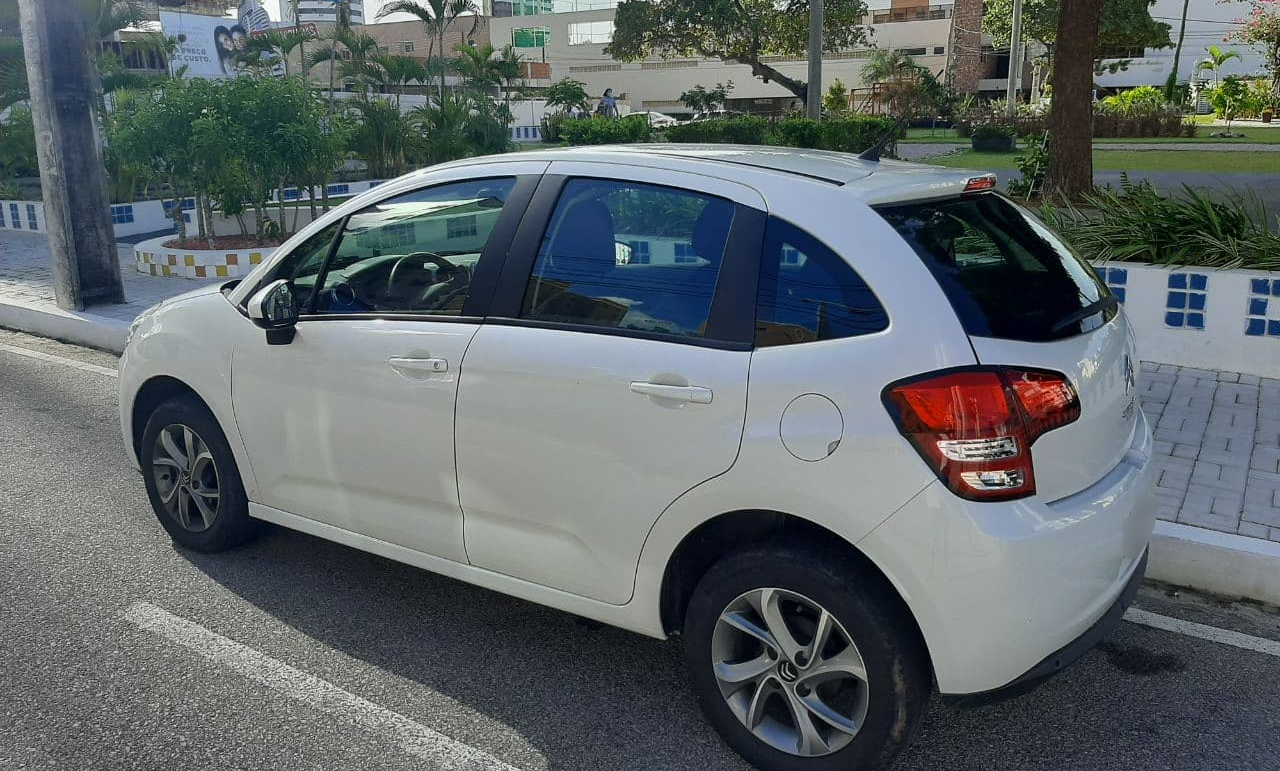}
\caption{Vehicle used in the second campaign.}
\label{fig:carro_2_externo}
\end{figure}

\begin{figure}[H] 
\includegraphics[width=0.70\linewidth]{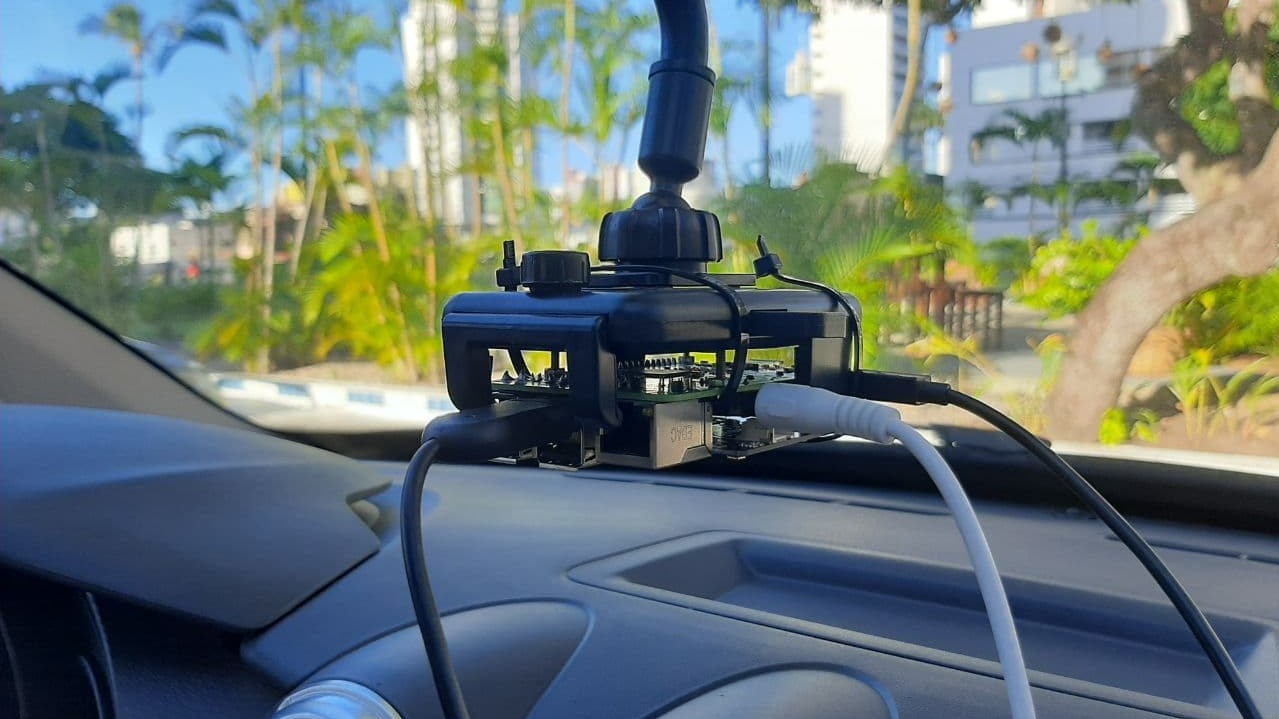}
\caption{Position of the measurement setup above the panel ({second} campaign).}
\label{fig:carro1}
\end{figure}

The car selected for the first campaign  was a C4 Lounge (Figure~\ref{fig:carro_1_externo}), and the one used in the second campaign was a C3 (Figure~\ref{fig:carro_2_externo}), which are both from Citroën and with automatic transmission. The boards were positioned above the cars' panels, as pictured in Figure~\ref{fig:carro1}. This position was chosen to mimic that of microphones in vehicle multimedia systems.

\vspace{-6pt}

 \vspace{-6pt}

\section{Statistical Methods}
\label{sec:methods} 

\begin{figure}[H] 
\includegraphics[width=.95\linewidth]{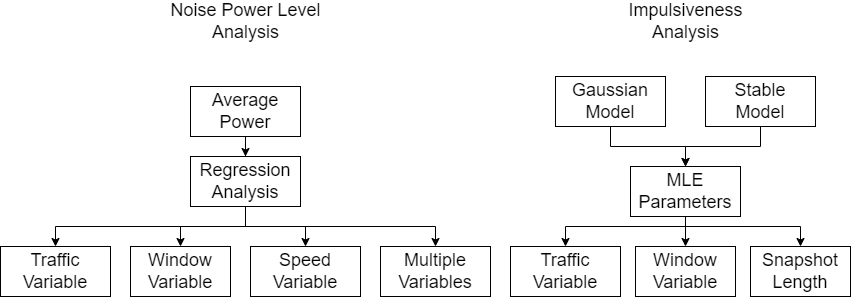}
\caption{Methodology flow chart.}
\label{fig:methodology}
\end{figure}

Our goal is to understand how the selected variables affect the characteristics of the acoustic noise inside the vehicle, namely the noise levels and the impulsiveness. To evaluate the first, the average power of each noise sample was computed, and statistical analysis was performed to assess the most relevant variables. To evaluate impulsiveness, we fit the noise data to the AWGN and alpha-stable models and compare their performance and how each variable affects the distribution of the models' parameters, as described in Figure~\ref{fig:methodology}.

\subsection{Average Power}


The average power of the measurements is computed for each individual acquisition as follows: 
\begin{equation}
\label{eq:power}
    P_{dBV} = 20 \cdot \text{log}\left(\frac{1}{N}\sum_{n=1}^{N} {{x^2[n]}} \right) \text{,}
\end{equation} 
\noindent where $N$ is the length of sampling, and $x(n)$ is the voltage signal from the microphone.

Usually, the power of acoustic signals and noise is measured using specific tools, such as a Sound Pressure Level (SPL) meter. We measured some sampling points with our setup and an MSL-1352C (Minipa) SPL meter to verify the measured power levels, whose specifications are described in Table~\ref{Tab:SPL_specs}. The meter was set to use A-weighting, slow response (1 s)\, and a range of 30 to 130~dB, as instructed by the meter's manual for measurement of the SPL of an oscillating noise. The SPL was positioned near the ReSpeaker setup, in the car's panel, and acquired data simultaneously as the microphone in the ReSpeaker. Unlike the ReSpeaker acquisition, the SPL measurements were manual.

\begin{figure}[H] 
    \includegraphics[width=0.9\linewidth]{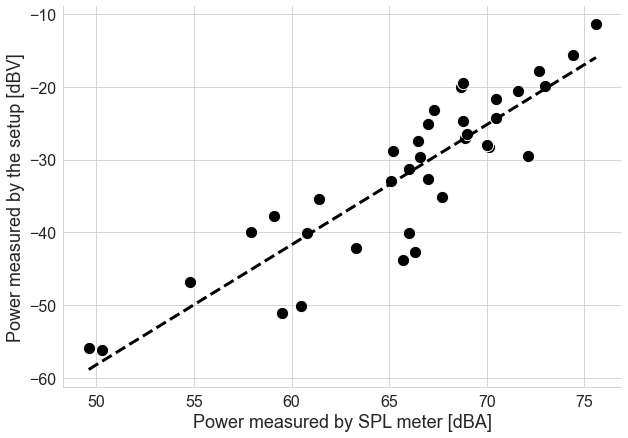}
    \caption{{Comparison between} power levels measured by the setup and the SPL meter.}
    \label{fig:decibelimetro} 
\end{figure}

Figure~\ref{fig:decibelimetro} compares the power levels measured by the instruments.  Despite the instrumentation errors present in this scheme, such as the positioning of the instruments and the reading of the SPL meter by a human, we can observe a linear relationship between the values measured by each instrument. 

\begin{table}[H]
\centering
\caption{Description of the MSL-1352C SPL specifications.}
\label{Tab:SPL_specs}
\setlength{\tabcolsep}{18.5mm}{\begin{tabular}{cc}
\toprule
\textbf{Specification} & \textbf{Value} \\ \midrule
Measurement range & 30$\sim$130 dB \\
Resolution & 0.1 dB \\
Frequency response & 31.5$\sim$8.5 kHz \\
Precision (94 dB/1 kHz) & $\pm$1.5 dB \\
Data logger capacity & 4422 samples \\ \bottomrule
\end{tabular}
}
\end{table}

\subsection{Regression Analysis}\label{sec:regression_analysis}

Each noise sample has four features: average noise power, windows positions, traffic conditions, and speed. Some of them are numeric in nature (power and speed), while the others are categorical. The \textit{Window} variable was encoded with 0s and 1s due to being a binary variable. The \textit{Traffic} variable is encoded in descending order of severeness, where \textit{Green} corresponds to 3, and \textit{Black} corresponds to 0.

We employ tools from descriptive and inferential statistics to analyze how the average power is related to the other variables. Boxplots are used to compare the power levels between different categories and histograms and density curves to visualize how it is distributed. 

Next, we create linear regression models for the continuous and \textit{Traffic} variables. The objective of the models is to highlight the relations between the selected variables and the power levels measured inside the car. The models obtained cannot be considered noise models for the acoustic noise in this scenario, as there are not enough samples for this type of characterization nor are sufficient measurements conditions being considered (such as multiple models of vehicle, for instance). Nevertheless, the use of linear regression models allows us to visualize the relations between the variables. Even if this relation is only approximately linear, the models can identify and quantify the effect of input on the output of a system.

Consider a set of $N$ pair of observations $(x_i,y_i) = (x_1,y_1),(x_2,y_2),...,(x_N,y_N)$; the simple linear regression model (with a single independent variable) for this set of observations is given by:

\begin{equation}
    y_i = \beta_0 + \beta_1 x_i + \epsilon_i, \qquad i= 1,2,...,N,
\end{equation}

\noindent with $y$ as the dependent or response variable, and $x$ as the independent, explanatory, or regression variable, and $\beta_0$ and $\beta_1$ as the regression or model coefficients~\cite{Stigler1981}. The model coefficients are estimated using the Ordinary Least Squares (OLS) method~\cite{Stigler1981}.

Three {Goodness of Fit} (GoF) metrics~\cite{devore_2011} are used to compare the results of the models: Mean Squared Error (MSE), coefficient of determination (R\textsuperscript{2}), and F-statistic. We use the logistic regression model~\cite{devore_2011} for categorical data and the McFadden Pseudo-R\textsuperscript{2}~\cite{scott2000} coefficient as a GoF metric to compare the models.

\subsection{Impulsive Noise and Alpha-Stable Model}\label{sec:impulsive_noise}

One of the most ubiquitous noise models in communications systems is the AWGN model, which is based on the Gaussian distribution. The use of the Gaussian distribution is motivated by the Central Limit Theorem, which states that the distribution of the sample mean of $N$ independent and identically distributed (i.d.d.) random variables with finite variance converges to a Gaussian distribution as $N\to\infty$~\cite{devore_2011}. Thus, the distribution is suited for modeling the cumulative effect of many independent noise sources. 

Despite its importance, the Gaussian model is not always the best choice to represent the noise in a communication channel~\cite{Nikias_1994}. Impulsive phenomena, when the noise varies subtly and greatly from the mean in a short period, can jeopardize the performance of solutions and strategies based on the traditional Gaussian approach~\cite{Yardimici_1998,Georgiou1999,Georgiou_2006}. Impulsive noise is present in several scenarios of communication systems, such as powerline communications~\cite{Bai2018}, OFDM in wireless networks~\cite{Landa_2020}, and sensor networks~\cite{Alam_2018}. Unlike the Gaussian, the alpha-stable distribution can have infinite variance, so it better represents data with heavy tails~\cite{shao1993,Nikias_1994}.

The family of stable distributions, also known as Lévy's alpha-stable distribution, comprises a class of distributions that satisfy the stability property~\cite{Nolan2020}: a random variable $X$ is said to be stable if for two independent instances $X_1$ and $X_2$ of $X$ and for any positive constants $a$ and $b$, the variable $aX_1 + bX_2$ has the same distribution that the variable $cX + d$, for $c>0$ and $d \in \mathbb{R}$.

In other words, a linear combination of i.d.d. stable variables will have the same distribution, except possibly for the location and scale parameters. Another essential property of the stable distributions is that they generalize the Central Limit Theorem. Relaxing the constraint of finite variance, the limit of the sum of i.d.d. random variables tends to a stable distribution. The Gaussian distribution and the traditional Central Limit Theorem are special cases when the variances of the random variables are finite~\cite{Nolan2020}.

The alpha-stable distribution has several different parametrizations. As found in recent literature~\cite{Nolan2020}, the most common form is to describe the distribution by its characteristic function $\phi (t)$:

\begin{equation}
    \phi (t) = \exp(j\delta t - \gamma |t|^{\alpha}[1+j\beta \text{sign}(t) \omega(t,\alpha)])
\end{equation} 
\noindent and 
\begin{equation}
    \omega(t,\alpha) = 
        \left\{
        \begin{array}{l}
             \tan{\frac{\alpha \pi}{2}}, \text{if } \alpha \neq 1 \\
             \frac{2}{\pi}\log{|t|},     \text{if } \alpha = 1
        \end{array}
        \right.
\end{equation}

\begin{equation}
    \text{sign}(t) = 
        \left\{
        \begin{array}{l}
             1, \text{if } t > 0 \\
             0, \text{if } t = 0 \\
             -1, \text{if } t < 0
        \end{array}.
        \right.
\end{equation}

The four parameters in the alpha-stable distribution are as follows:

\begin{itemize}[align=parleft,leftmargin=*,labelsep=3mm]
    \item $\alpha$, the characteristic exponent, satisfying $0 < \alpha \leq 2$. It is the main shape parameter of the distribution, describing the tails of the distribution. Smaller values of $\alpha$ indicate a heavier tail, meaning a higher probability of extreme events. Conversely, values approaching $2$ indicate a behavior closer to that of a Gaussian distribution. When $\alpha = 2$, it is equivalent to a Gaussian distribution; 
    \item $\beta$, the skewness parameter, is limited to $\beta \in [-1,1]$. It controls the skewness of the distribution. For $\beta = 0$, the distribution is symmetric. If $\beta > 0$, then the distribution is right-skewed. If $\beta < 0$, then the distribution is left-skewed; 
    \item $\gamma$, the scale parameter, which is always a positive number ($\gamma >0$).
    This parameter behaves similarly to the variance in the Gaussian distribution. It determines the dispersion around the location parameter. It should be noted that the variance of an alpha-stable variable is only defined for $\alpha = 2$; 
    \item $\delta$, the location parameter, which shifts the distribution to the left or to the right by an amount $\delta \in \mathbb{R}$.  
\end{itemize}

Lastly, we highlight that the distributions with $\beta = 0$ and $\delta = 0$ form a particular family of symmetric stable distributions known as Symmetric $\alpha$-Stable ($S\alpha S$). These distributions share many characteristics with the Gaussian distribution. Both are continuous, unimodal, and bell-shaped distributions. The main difference is in the decay of the tails: the Gaussian curve has an exponential decay, while the $S\alpha S$ has an algebraic one~\cite{shao1993}. These properties make the $S\alpha S$ model a common choice to model problems in signal processing where the distribution is similar to the Gaussian but with heavier tails~\cite{shao1993,Georgiou1999,BELKACEMI_2007}.

To evaluate the degree of impulsiveness in the interior vehicle noise, as well as to compare the performance of the two models, we estimate the parameters of a Gaussian and a stable distribution fitted to all the collected samples. The fitting of the noise samples to the models is obtained with Maximum Likelihood Estimation (MLE)~\cite{scharf_1991}.

The application of MLE for the Gaussian case is straightforward. In the case of the stable distributions, for which no closed expression for the probability density function exists, the MLE must be found with numerical methods and optimizations routines~\cite{Nolan2020}. In this work, we computed the MLE using MATLAB, which bases its implementation on the works of John P. Nolan~\cite{Nolan1997,Nolan2001}. To obtain a starting point to the optimization routine, MATLAB uses the method described in~\cite{McCulloch1986}. In this approach, the four parameters are derived in terms of five quantiles of the data. Although the accuracy of this method is inferior, its low computational cost makes it convenient to provide a starting point for other estimation techniques.  

\section{Results and Discussions}
\label{sec:results}

Table~\ref{tab:numero_amostras} shows the number of samples collected in each measurement campaign as well as the encoding used for the variables \textit{Traffic} and \textit{Window}. There is a balanced amount of samples for the two window position situations in both campaigns and in Campaign 1 for the traffic categories. In the case of Campaign 2, there is a higher amount of \textit{Black} category samples. The change in traffic patterns imposed by the COVID-19 pandemic sanitary measures made it difficult to obtain samples in this category, as traffic became lighter than usual for some of the roads where traffic jams usually happen. 
However, it was possible to obtain an equivalent number of samples for the \textit{Black} category to that of Campaign 1.

In this section, we analyze the results using the methodology described in Section~\ref{sec:methods} and illustrated in Figure~\ref{fig:methodology}. The analyses were performed using acoustic signals measured with setup and constraints described in Section~\ref{sec:measurement_setup} for each condition described in Table~\ref{Tab:variaveis_ambiente}. Finally, the evaluation metrics used are described in Sections~\ref{sec:regression_analysis} and~\ref{sec:impulsive_noise}.

\begin{table}[H] 
\caption{Number of samples collected in each measurement campaign by traffic condition and position of windows. The last row presents the encoding used to model the variables \textit{Traffic} and~\textit{Window}.}
\label{tab:numero_amostras}
\resizebox{\linewidth}{!}{%
\begin{tabular}{@{}cccccccc@{}}
\toprule
 & \multicolumn{2}{c}{\textbf{Window}} & \multicolumn{4}{c}{\textbf{Traffic}} & \multirow{2}{*}{\textbf{Total}} \\ \cmidrule(lr){2-7}
 & \textbf{Open} & \textbf{Closed} & \textbf{Black} & \textbf{Red} & \textbf{Orange} & \textbf{Green} &  \\ \midrule
\begin{tabular}[c]{@{}c@{}}Number of samples for\\ the first campaign\end{tabular} & 95 & 99 & 47 & 51 & 51 & 45 & 194 \\ \midrule
\begin{tabular}[c]{@{}c@{}}Number of samples for\\the second campaign\end{tabular} & 127 & 127 & 45 & 80 & 64 & 65 & 254 \\ \midrule
Encoding & 1 & 0 & 0 & 1 & 2 & 3 & - \\ \bottomrule 
\end{tabular}%
}
\end{table}

\subsection{Noise Power Level Analysis}

\begin{figure}[H] 
\includegraphics[width=.95\linewidth]{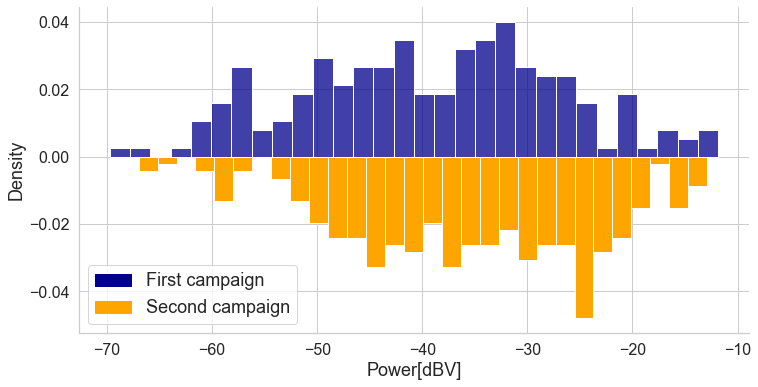}
\caption{\label{fig:histograma_both}Comparison of the distribution of the power levels measured in both campaigns. The histograms were normalized, the total area of each being unitary. The negative density in the second histogram is merely a consequence of its mirroring for illustrative purposes.}
\end{figure}

Figure~\ref{fig:histograma_both} shows the average power distributions of the collected noise samples for both campaigns. The negative density in the second histogram is merely a consequence of its mirroring for illustrative purposes. The range and distribution of values are similar for both campaigns, and a visual inspection indicates that most samples are concentrated in the center of the range. 


The histograms show that the campaigns have measurements with compatible power values. In the following subsections, an analysis of the average power in relation to the other three variables of the study (traffic, window position, and speed) is carried out individually, which is followed by an analysis of the three variables together.

\subsubsection{Traffic Analysis}
\label{subsec:traffic_analysis}
\begin{figure}[H] 
\includegraphics[width=.95\linewidth]{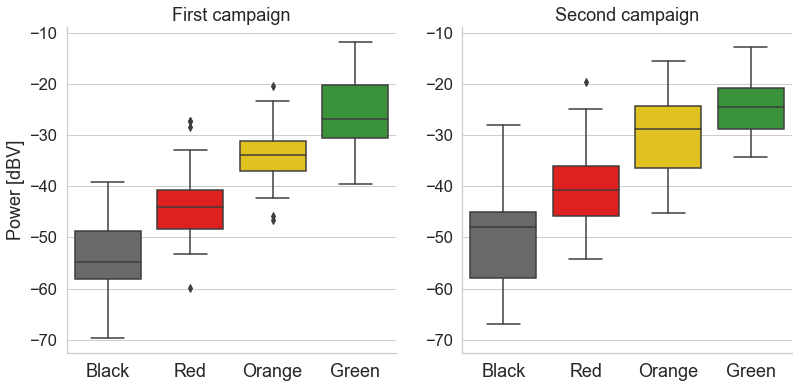}
\caption{Box diagrams of power levels by traffic category for both measurement campaigns.}
\label{fig:box_trafego_camp1}
\end{figure}

Figure~\ref{fig:box_trafego_camp1} presents the box diagrams of the power levels grouped by traffic conditions for Campaigns 1 and 2, respectively. For both cases, the boxes are ordered from \textit{Black} to \textit{Green} in ascending order of power, indicating that noise levels inside the car tend to increase as traffic becomes less severe. The main difference between the results of the campaigns is the greater variation in power in the first three traffic categories for the second campaign, which is illustrated in its taller boxes and lines. For example, there is a greater intersection between the power levels for the \textit{Yellow} and \textit{Green} categories in the second campaign, with samples from the \textit{Yellow} category reaching higher power levels.

Despite these differences, both results indicate the same trend toward higher noise levels associated with more fluid traffic conditions. These results suggest that there is some significant correlation between the two variables. To assess this relation, a linear regression model was built for each campaign in the form: 
\begin{equation}
    \textit{Traffic} \approx a_{0} + a_{1} \cdot \textit{power} \text{,}
\end{equation} 
where the intercept variable is $a_0$ and the coefficient of the explanatory variable (power level) is $a_1$. We chose to fit a linear model due to the ordered nature of the traffic data and the trend implied in Figure~\ref{fig:box_trafego_camp1}.

\begin{figure}[H]
\includegraphics[width=.95\linewidth]{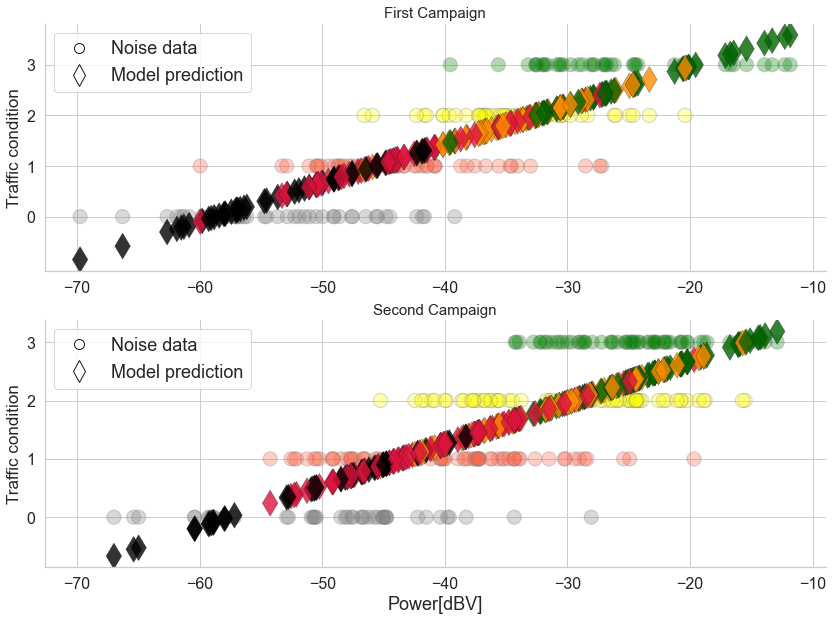}
\caption{\textit{Traffic} and predictions using the linear model for both campaigns. Colors represent the actual traffic conditions of each sample, which is {in accordance with Table~\ref{Tab:variaveis_trafego}.}}
\label{fig:model_traffic}
\end{figure}

The models obtained are shown in Figure~\ref{fig:model_traffic}. The circles represent the actual traffic condition associated with each sample, while the diamonds represent the traffic predicted by the model. Both models show that higher power levels imply a less severe traffic condition, which is in accordance with the behavior shown in the box diagrams. The range of predictions for each traffic category is centered around the correct value for the \textit{Traffic} variable, although some variation causes overlap between the categories. For instance, this can be seen in the red diamonds centered around $\textit{Traffic} = 1$.

A comparison between the two models in Figure~\ref{fig:model_traffic} highlights the greater variability of the data in Campaign 2, which can also be seen in the box diagrams. This is also reflected in the GoF metrics listed in Table~\ref{tab:traffic}. The first model has a higher value for R\textsuperscript{2} and for the F-Statistic and a lower value for MSE, confirming its better performance. In fact, the results indicate that 72\% of the power variability is explained by the variation in traffic in data of the first campaign. This suggests a strong relationship between the variables and that a large part of the observed indoor noise power is associated with the traffic level. The high F value and its low {\em p}-value confirm the significance of this relation. Although the results of the second model are inferior due to the higher dispersion of power in each traffic condition, they too indicate a significant association between power and traffic, with 61\% of the variation explained by the model. In both campaigns, the MSE has a low value. However, due to the categorical nature and scale of the traffic data, the MSE is not an adequate GoF metric for the \textit{Trafic} models.

\begin{table}[H] 
\caption{Regression coefficients (with 95\% confidence interval) and GoF metrics for the power and traffic~models.}
\label{tab:traffic}
\resizebox{\linewidth}{!}{%
\begin{tabular}{@{}ccccccc@{}}
\toprule
 & \multicolumn{2}{c}{\textbf{Regression Coefficients}} & \multicolumn{4}{c}{\textbf{Goodness of Fit}} \\ \cmidrule(l){2-7} 
 & \boldmath{$a_0$} & \boldmath{$a_1$} & \textbf{MSE} &\textbf{ R²} &\textbf{ F-Value} & \textbf{Prob (F)} \\ \midrule
1st campaign & 4.4942 (4.216--4.772) & 0.0766 (0.070--0.083) & 0.337 & 0.72 & 498.26 & \num{3.03e-55} \\ \midrule
2nd campaign & 4.1012 (3.841--4.361) & 0.0712 (0.064--0.078) & 0.431 & 0.61 & 401.96 & \num{4.19e-54} \\ \bottomrule
\end{tabular}%
}
\end{table}

\subsubsection{Window Analysis}

\begin{figure}[H] 
\includegraphics[width=0.85\linewidth]{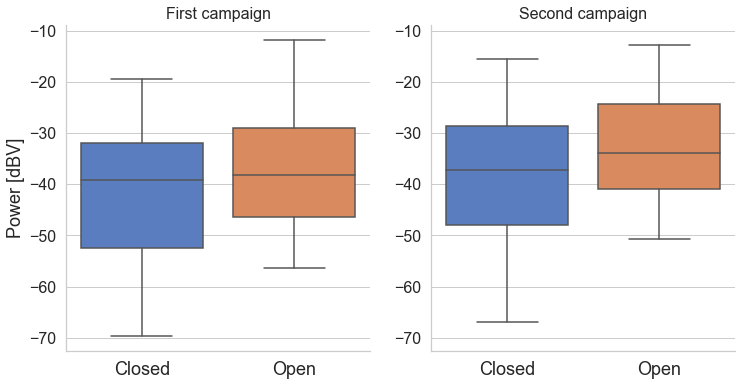}
\caption{Box diagrams of the average noise power of the samples grouped by the variable \textit{Window}.}
\label{fig:box_janela}
\end{figure}

Figure~\ref{fig:box_janela} presents the box diagrams for the power levels grouped by the position of the car windows for both campaigns. Comparing the campaigns, 
the power levels for the second are slightly 
higher than the first. Unlike the \textit{Traffic} variable, the layout of the samples is visually very similar between the two categories. In both campaigns, power levels tend to be higher when the windows are open, which is expected, as there is more coupling of outside noise inside the car. However, there is a significant overlap in values between the two categories. For Campaign 1, only 8.42\% of the measurements in the \textit{Open} group have a power greater than the maximum power in the \textit{Closed} group, while for Campaign 2, the percentage is 4.72\%.

This implies few distinctions in power values when the car windows are open or closed. This result goes against expectations, as the qualitative difference is significant when perceived by a passenger or when listening to the recordings of this experiment. However, this sensorily-perceived difference does not manifest itself in an expressive difference in the average power level received by a microphone located close to the vehicle's panel, which can be advantageous for voice command applications. 

\begin{figure}[H] 
\includegraphics[width=0.85\linewidth]{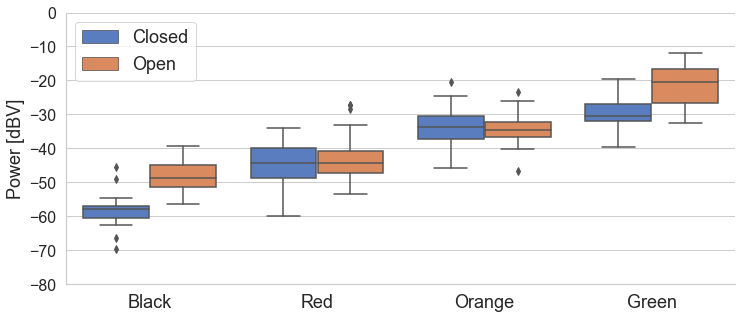}
\caption{Power levels box diagrams grouped by both \textit{Traffic} and \textit{Window} variables (fist campaign).}
\label{fig:box_janela_trafego_camp1}
\end{figure}

\begin{figure}[H] 
\includegraphics[width=0.85\linewidth]{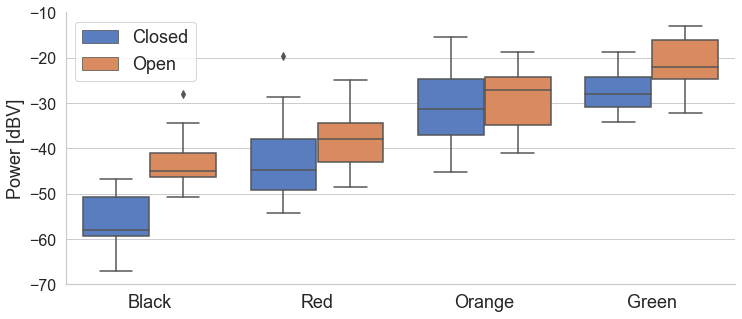}
\caption{Power levels box diagrams grouped by both \textit{Traffic} and \textit{Window} variables (second~campaign).}
\label{fig:box_janela_trafego_camp2}
\end{figure}

To verify whether this observation has any bias in relation to the \textit{Traffic} variable, Figures~\ref{fig:box_janela_trafego_camp1} and~\ref{fig:box_janela_trafego_camp2} present the box diagrams of the samples grouped by traffic and window position for both sets of measurements. Once more, the results for the two campaigns are in agreement. The most significant difference between the two windows positions occurs in the \textit{Black} category when the vehicle is stopped in a traffic jam.

We speculate that in this case, the absence of movement of the car makes external noises predominate, and the position of the windows becomes more significant than in other scenarios. As it gains speed, the noise generated by the vehicle becomes more relevant, so that for \textit{Red} or \textit{Yellow} traffic, the difference between the power levels is small between both states of the windows. Finally, when the vehicle reaches a higher speed (category \textit{Green}), there is again a noise trend of higher power values when the windows are open. It is speculated that at these speeds, the noise produced by the wind, thanks to the car's movement, has a more significant contribution.

\vspace{-6pt}

\begin{figure} [H]
    \includegraphics[width=0.9\linewidth]{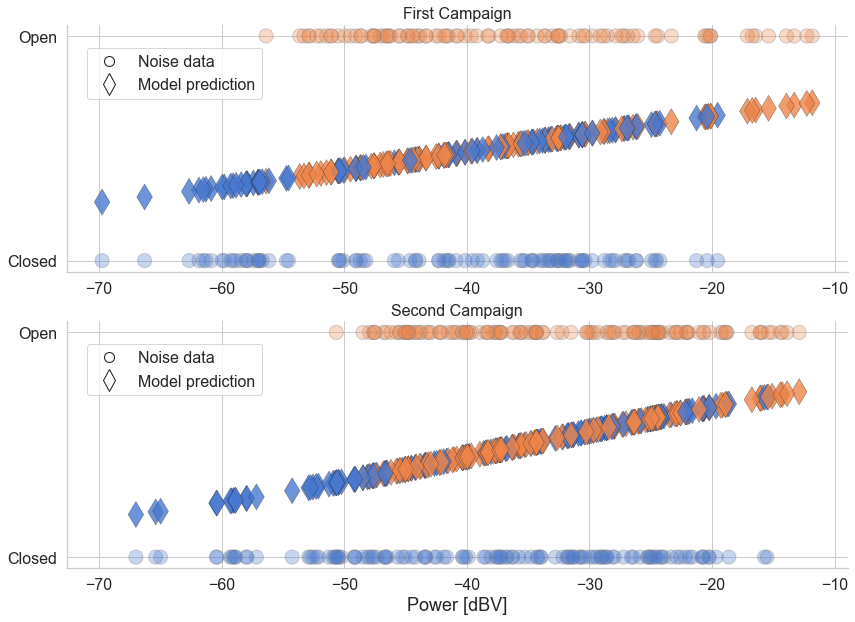}
    \caption{\textit{Window} data and predictions using the logistic model for both campaigns.}
    \label{fig:model_window}
\end{figure}

Then, the overall effect observed is of slightly higher noise power when the windows are open. As there is little distinction between the power levels of the groups, it is expected that a model that takes into account only the average power of the samples will be unable to represent the data well. Given the categorical and binary nature of the variable in question, two logistic models are obtained in the following form: 
\begin{equation}
    \textit{window} \sim \frac{e^{b_{0} + b_{1} \cdot \textit{power}}}{1 + e^{b_{0} + b_{1} \cdot \textit{power}}
     } \text{ ,}
\end{equation}
\noindent where $b_0$ and $b_1$ are the model coefficients. Table~\ref{Tab:gof_janela_camp1} presents the coefficients and GoF metrics, while Figure~\ref{fig:model_window} shows the models obtained. Visually, it is clear that there is little differentiation between the categories. If the models were used for a classification task, the accuracy would be substandard. The Pseudo-R\textsuperscript{2} values of both models are low and close to each other. These results suggest a weak influence of the window position on the measured internal power. It is important to emphasize that this result considers all traffic categories. Figures~\ref{fig:box_janela_trafego_camp1} and~\ref{fig:box_janela_trafego_camp2} show that the distinction between power levels is greater for extreme traffic categories. Finally, in contrast to the \textit{Traffic} variable, the performances of the \textit{Window} models are quite similar for the two measurement campaigns.

\vspace{-6pt}

\begin{table}[H] 
\caption{Regression coefficients (with 95\% confidence interval) and GoF metrics for the power and window models.}
\label{Tab:gof_janela_camp1}
\resizebox{\linewidth}{!}{%
\begin{tabular}{@{}ccllcllclll@{}}
\toprule
 & \multicolumn{6}{c}{\textbf{Regression Coefficients}} & \multicolumn{4}{c}{\textbf{Goodness of Fit}} \\ \cmidrule(l){2-11} 
 & \multicolumn{3}{c}{\boldmath{$b_0$}} & \multicolumn{3}{c}{\boldmath{$b_1$}} & \multicolumn{4}{c}{\textbf{Pseudo-R²}} \\ \midrule
1st campaign & \multicolumn{3}{c}{1.2482 (0.254--2.242)} & \multicolumn{3}{c}{0.0329 (0.009--0.057)} & \multicolumn{4}{c}{0.0274} \\ \midrule
2nd campaign & \multicolumn{3}{c}{1.6167 (0.771--2.463)} & \multicolumn{3}{c}{0.0458 (0.023--0.069} & \multicolumn{4}{c}{0.04704} \\ \bottomrule
\end{tabular}%
}
\end{table}

\subsubsection{Speed Analysis}
\label{subsec:speed_analysis}

\begin{figure}[H] 
\includegraphics[width=0.9\linewidth]{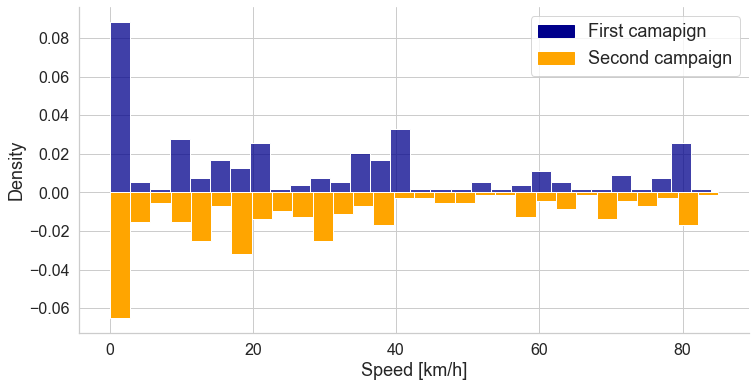}
\caption{Comparison of the speed values for the samples between measurement campaigns. The histograms were normalized, the total area of each being unitary. The negative density in the second histogram is merely a consequence of its mirroring for illustrative purposes.}
\label{fig:hist_speed_both}
\end{figure}

The histogram in Figure~\ref{fig:hist_speed_both} shows the speed distribution of the measurements from both campaigns. There is a larger number of measurements for zero velocity. These points correspond to the \textit{Black} traffic category, when the car is stationary or at a very low speed due to traffic jams. The speed of the car during measurement is linked to the traffic condition at the time of measurement (Table~\ref{Tab:variaveis_trafego}). Therefore, there is a greater concentration of measurements in the ranges between 0 and 20 km/h (\textit{Red}), and 20 and 40 km/h (\textit{Orange}), when compared to the longer range of 40 to 80 km/h (\textit{Green}). In both campaigns, we sought to measure at different speeds to take into account the entire speed range of each traffic~level.

\begin{figure}[H] 
\includegraphics[width=0.95\linewidth]{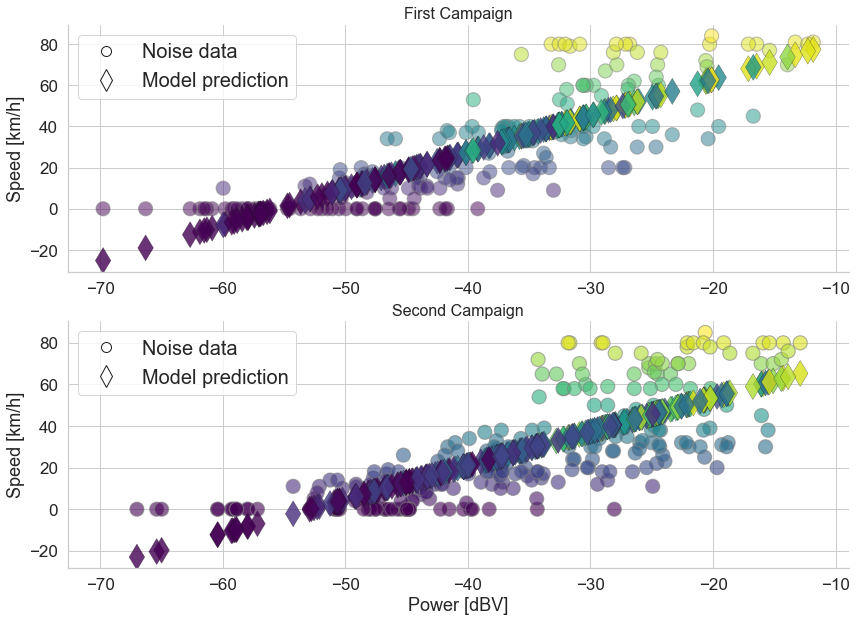}
\caption{Speed data and predictions using the linear models.}
\label{fig:model_speed_camp1}
\end{figure}

Of the three variables analyzed, speed is the only numerical one in nature. Therefore, a linear regression model is obtained in the form: 
\begin{equation}
    \textit{speed} \approx c_{0} + c_{1} \cdot \textit{power} \text{,}
\end{equation} 
\noindent where $c_0$ and $c_1$ are the regression coefficients. Figure~\ref{fig:model_speed_camp1} shows the predictions of the models and the actual speed values, while Table~\ref{tab:speed} lists the GoF metrics. The figures indicate that a higher speed implies higher noise levels, which is expected, as higher speeds result in more engine noise and more vibrations in other parts of the vehicle. 

\begin{table}[H] 
\caption{Regression coefficients (with 95\% confidence interval) and GoF metrics for the power and speed~models.}
\label{tab:speed}
\resizebox{\linewidth}{!}{%
\begin{tabular}{@{}ccccccc@{}}
\toprule
 & \multicolumn{2}{c}{\textbf{Regression Coefficients}} & \multicolumn{4}{c}{\textbf{Goodness of Fit}} \\ \cmidrule(l){2-7} 
 & \boldmath{$c_0$} & \boldmath{$c_1$} & \textbf{MSE} & \textbf{R²} & \textbf{F-Value} & \textbf{Prob (F)} \\ \midrule
1st campaign & 98.37 (91.23--105.51) & 1.77 (1.60--1.94) & 221.65 & 0.68 & 404.92 & \num{3.56e-49} \\ \midrule
2nd campaign & 86.33 (79.81--92.85) & 1.63 (1.46--1.81) & 270.76 & 0.57 & 336.34 & \num{2.66e-48} \\ \bottomrule
\end{tabular}%
}
\end{table}

Visually, both models show a good fit to the data. This is also confirmed by the GoF metrics. The ~R\textsuperscript{2} value indicates that approximately 68\% and 57\% of the variation in power is explained by the variation in speed in models 1 and 2, respectively. The high F-value and the low {\em p}-value confirm that the relation between the variation in power and the variation in speed expressed by the models is unlikely to result from chance. 

Similar to what was discussed for the \textit{Traffic} models, the dispersion of power levels is greater in the Campaign 2 samples. The circles in Figure~\ref{fig:model_speed_camp1} illustrate the greater variability of power levels in the second campaign 
in the speed range from 0 to 40 km/h. This interval is in accordance with the greater variability in power values observed when comparing the box diagrams shown in Figure~\ref{fig:box_trafego_camp1} for the \textit{Black}, \textit{Red}, and \textit{Yellow} conditions.

The models obtained for the variables \textit{Traffic} and \textit{Speed} are similar in appearance and GoF metrics. On the other hand, the difference between the two measurement groups is smaller for the \textit{Window} variable. This set of results suggests that the average power level measured inside the vehicle is generally more influenced by the car's movement, which depends on its speed and traffic conditions, than by external noise sources. Furthermore, the similarity between the adjustments of the variables \textit{Traffic} and \textit{Speed} suggests a strong correlation between the two, arising from the way traffic levels are defined in Table~\ref{Tab:variaveis_trafego}. 

\subsubsection{Multiple Variable Analysis}

\begin{figure}[H] 
\includegraphics[width=0.93\linewidth]{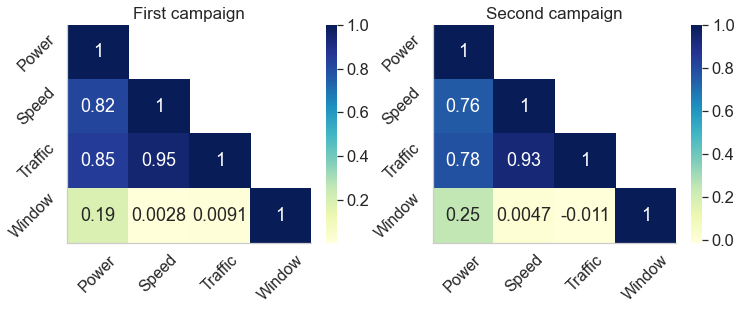}
\caption{Correlation matrix of the dataset for both measurement campaigns.}
\label{fig:corr}
\end{figure}

The \textit{Window} variable analysis demonstrates that the windows have a weak influence on the vehicle noise level. Nonetheless, the traffic and speed variables contribute significantly to the vehicle interior noise in the traffic and speed analyses. One way to check the strength of the relationship between variables and noise level is to calculate their cross-correlation. Figure~\ref{fig:corr} presents the correlation matrix for the two datasets. In both, noise power has a high correlation with traffic and speed and a low correlation with window position.


\begin{figure}[H] 
\includegraphics[width=0.9\linewidth]{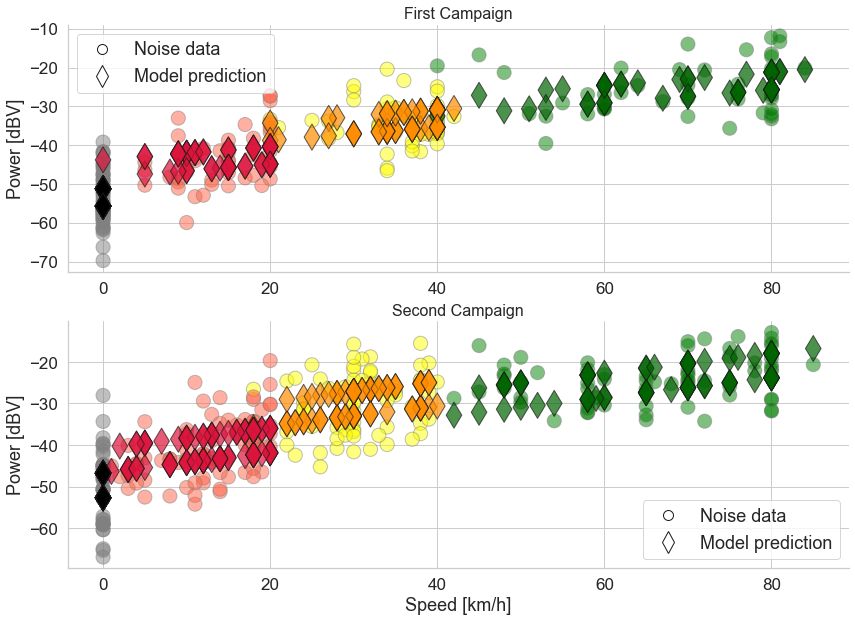}
\caption{Power data grouped by traffic conditions, and predictions using the linear models with \textit{Speed}, \textit{Window}, and \textit{Traffic} as the explanatory variables. {Colors represent the actual traffic conditions of each sample, which are in accordance with Table~\ref{Tab:variaveis_trafego}.}
}

\label{fig:model_all}
\end{figure}

Figure~\ref{fig:corr} also shows a high correlation between \textit{Traffic} and \textit{Speed}. A high correlation is to be expected due to how the traffic conditions are defined by Google (Table~\ref{sec:campaign_conditions}). In the context of statistical modeling, the variables convey roughly the same information about the noise power inside the vehicle. To better illustrate this redundancy, two models of the average noise power are created using the other characteristics as independent variables (Figure~\ref{fig:model_all} and Table~\ref{tab:all}). The models are a linear regression with categorical and numerical variables, in the form 
\begin{equation}
\label{eq:model_all}
\begin{split}
        \textit{power} \approx  &  d_{0} + d_{1} \cdot \textit{speed} + d_{2} \cdot \textit{traffic}_{\textit{red}} + d_{3} \cdot \textit{traffic}_{\textit{orange}} + d_{4} \cdot \textit{traffic}_{\textit{green}} + \\
        & + d_{5} \cdot \textit{window}_{\textit{open}} 
    \text{ ,}
\end{split}
\end{equation} 
\noindent where $d_0$ is the intercept; $d_1$ is the coefficient of \textit{Speed}; $d_2$, $d_3$, and $d_4$ are the coefficients added when the traffic condition is red, yellow, or green, respectively; and $d_5$ is the coefficient added when windows are open.

\vspace{-6pt}

\begin{table}[H] 
\caption{Model coefficients and GoF metrics for the power model vs. other variables.}
\label{tab:all}
\resizebox{\linewidth}{!}{%
\begin{tabular}{@{}ccccccccccc@{}}
\toprule
 & \multicolumn{6}{c}{\textbf{Regression Coefficients}} & \multicolumn{4}{c}{\textbf{Goodness of Fit}} \\ \cmidrule(l){2-11} 
 & \boldmath{$d_0$ }& \boldmath{$d_1$} & \boldmath{$d_2$} & \boldmath{$d_3$} & \boldmath{$d_4$} & \boldmath{$d_5$} & \textbf{MSE} & \textbf{R²} &\textbf{ F-Value} & \textbf{Prob (F)} \\ \midrule
1st campaign & $-$55.72 & 7.44 & 13.56 & 16.45 & 4.492 & 0.170 & 35.63 & 0.766 & 123.01 & \num{2.50e-57} \\ \midrule
2nd campaign & $-$52.69 & 6.09 & 12.61 & 9.92 & 5.87 & 0.27 & 39.59 & 0.713 & 123.086 & \num{4.21e-65} \\ \bottomrule
\end{tabular}%
}
\end{table}

Coefficients $d_2$, $d_3$, and $d_4$ from Equation (\ref{eq:model_all}) determine a base power level for each traffic category. This is illustrated in the graphs above: the predictions are grouped  according to their traffic condition. Each group has two straight lines corresponding to the two possible states of the \textit{Window} variable. These lines are close together for all traffic conditions, with a small difference between the two. This difference is compatible with the box diagrams discussed previously, which indicate that the power tends to be slightly higher when the windows are open. In fact, the variable \textit{Speed}, related to the slope of the eight line functions, is what determines the power in each group of traffic conditions.

The high values of R\textsuperscript{2} in Table~\ref{tab:all} indicate that most of the variation in noise power is accounted for by the models. However, a comparison with the GoF metrics in Sections~\ref{subsec:speed_analysis} and~\ref{subsec:traffic_analysis} show that although the R\textsuperscript{2} increases greatly in the more complex model, the F-value decreases with the addition of the other two variables. The \textit{Traffic} variable contributes little to the model due to its redundancy with \textit{Speed}, while \textit{Window} has almost no relation to the response variable. We conclude that either the traffic categories or the speed can be used as an explanatory variable for the noise power inside vehicles due to the way traffic was defined in this work. 

\subsection{Impulsiveness Evaluation}

For this section, the alpha-stable and Gaussian distributions were fitted to each noise sample using MLE. To calculate the fitting error for the two distributions, a histogram was created based on the empirical cumulative distribution of each measurement. Then, the Root Mean Squared Error (RMSE) was calculated between a probability density curve with the parameters estimated by the MLE and the data histogram.

As in the first section, this analysis is split between the variable \textit{Traffic} and the variable \textit{Window}. The variable \textit{Speed} is omitted for clarity, since it would have redundant results with the variable \textit{Traffic}. For the stable distributions, we assume an $S\alpha S$ model and estimate only the $\alpha$ and $\gamma$ parameters. 

\subsubsection{Traffic Analysis}

\begin{figure}[H] 
    \includegraphics[width=.95\linewidth]{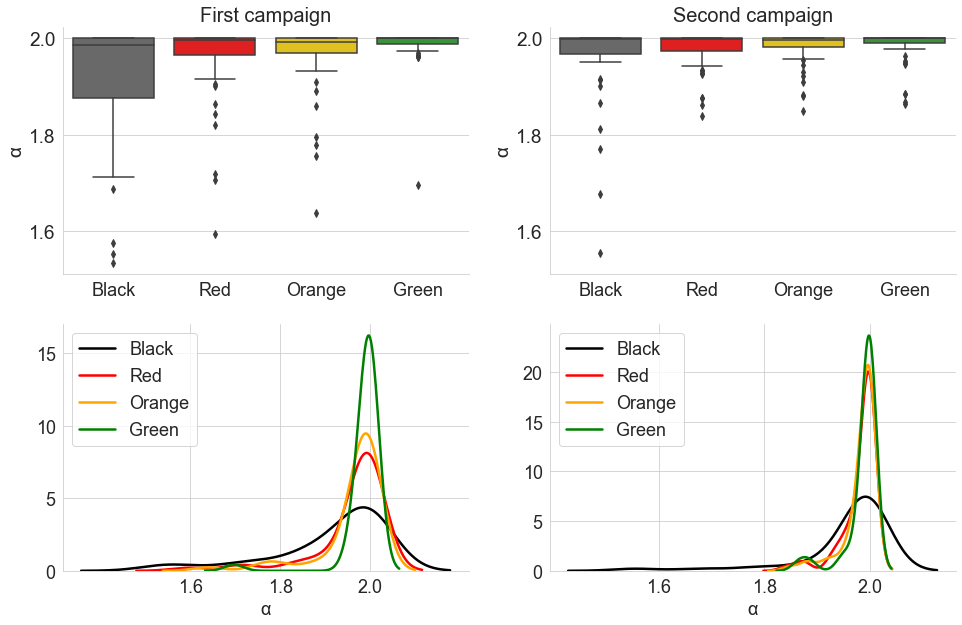}
    \caption{Distribution of the estimated parameter $\alpha$, from the alpha-stable distribution, grouped by the traffic conditions. The KDE curves were obtained using a Gaussian kernel.}
    \label{fig:alpha_trafego}
\end{figure}

\begin{figure}[H] 
    \includegraphics[width=.95\linewidth]{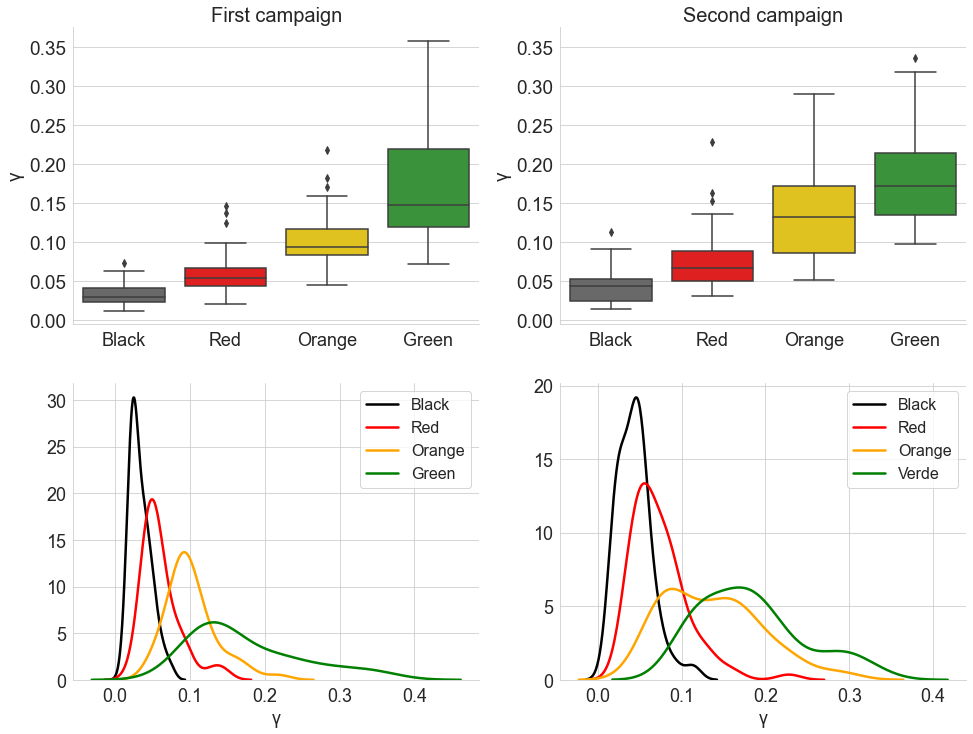}
    \caption{Distribution of the estimated parameter $\gamma$, from the alpha-stable distribution, grouped by the traffic conditions. The KDE curves were obtained using a Gaussian kernel.}
    \label{fig:gamma_trafego}
\end{figure}

\begin{figure}[H] 
    \includegraphics[width=.95\linewidth]{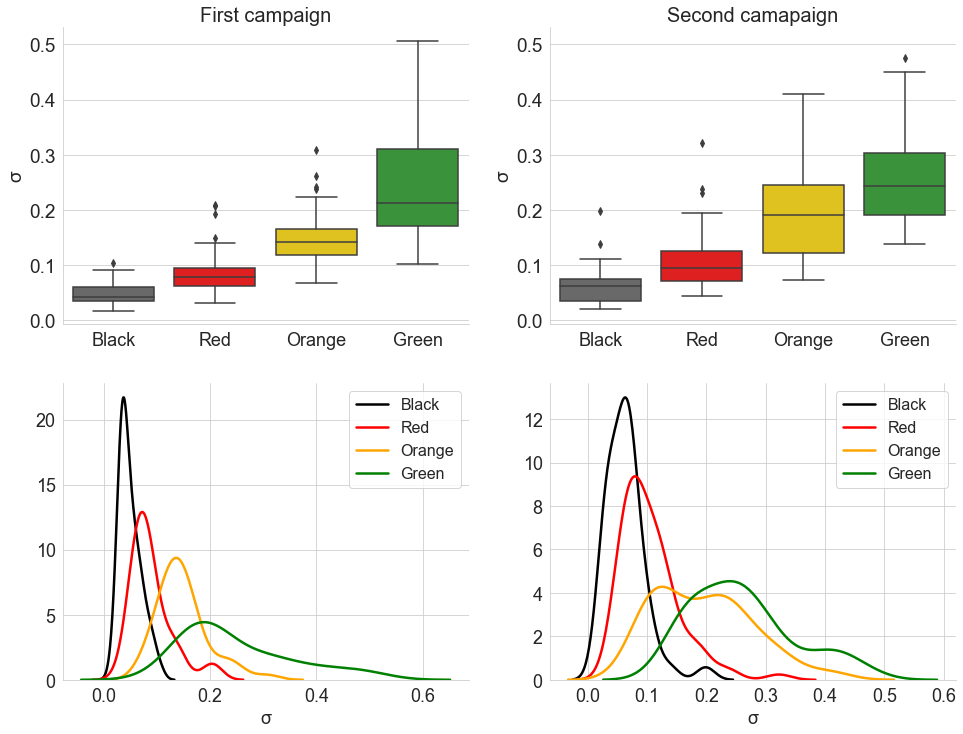}
    \caption{Distribution of the estimated parameter $\sigma$, from the Gaussian distribution, grouped by the traffic conditions. The KDE curves were obtained using a Gaussian kernel.}
    \label{fig:sigma_trafego}
\end{figure}

Figures~\ref{fig:alpha_trafego} and ~\ref{fig:gamma_trafego} show, respectively, the distribution of the estimated parameters $\alpha$~ and $\gamma$ of the alpha-stable distribution, which are grouped by traffic conditions. Figure~\ref{fig:sigma_trafego} shows the distribution of the parameter $\sigma$ of the Gaussian distribution. The parameter $\mu$ is close to zero for all measurements, indicating no \textit{offset} in the noise level.

In Figure~\ref{fig:alpha_trafego}, the estimated values in the categories \textit{Green}, \textit{Yellow}, and \textit{Red} are closer together and to $\alpha = 2$ in Campaign 2 than in Campaign 1. The first campaign also has more significant \textit{outliers}, which have metrics with $\alpha < 1.8$ across all traffic categories. This makes the results of Campaign 1 more dispersed. The difference is even better seen by comparing the Kernel Density Estimation (KDE) curves, which are more spread out around $\alpha = 2$ for the first campaign.

\vspace{-6pt}

However, the results of both campaigns are compatible with the behavior of $\alpha$ across the traffic categories. In general, the values are concentrated very close to $\alpha = 2$. The positions of the box diagrams and the fact that the KDE curves are centered close to this value demonstrate this fact. This suggests that most of the measured samples present behavior that can be well-represented by a Gaussian distribution, and, therefore, it can be said that they present low impulsiveness. However, a significant number of noise samples deviate from $\alpha = 2$ and can be said to present some degree of impulsiveness. As with the average power, this degree is ordered according to traffic categories: \textit{Green} presents the least impulsive behavior, with a greater proportion of samples close to $\alpha = 2$, while the \textit{Black} category has the widest range of values. The \textit{Red} and \textit{Yellow} categories have a similar distribution, being placed between the other two.

Combining this result with the one from Section~\ref{subsec:traffic_analysis}, we speculate that a lower degree of impulsiveness can be associated with a higher speed of the car. When in a \textit{Green} traffic situation, the continuous sound produced by the vehicle and the rolling noise are more dominant in the composition of the internal acoustic noise, so that it tends to have a characteristic closer to that of a Gaussian noise. Conversely, when a car stops in a traffic jam, sources such as other vehicles passing in the adjacent lanes, vehicles braking beside and behind the car, among other external sources, are more prevalent. These tend to be more transitory in nature, contributing to an increase in the observed impulsiveness.

Therefore, assuming an AWGN model for the vehicle's internal noise can be detrimental to algorithms and applications that suffer performance degradation in the presence of non-Gaussian noise. It is important to remember that in this study, sources of noise such as potholes and horns, the noise produced by magazines and local businesses, sounds generated by passengers, and events such as rain, among others, were disregarded. Therefore, the impulsive nature described above is optimistic, justifying research on more complex models that consider non-Gaussian behavior for the vehicular scenario.

The second estimated parameter of the alpha-stable model is the scale parameter $\gamma$~(Figure~\ref{fig:gamma_trafego}). An evaluation of the box diagrams shows that the distributions of values by traffic category are in ascending order. This result is very similar to what was discussed about the average power in Section~\ref{subsec:traffic_analysis}. Another similarity between the behavior of the average power and $\gamma$~ is the difference between campaigns. Again, the dispersion of values is greater in the second campaign, especially in the \textit{Yellow} category, which is similar to the comparison made between the campaigns in Figure~\ref{fig:box_trafego_camp1}.

In turn, the result for the estimation $\sigma$, in turn, shows great similarity with the distributions of $\gamma$. In fact, these parameters have a similar nature, being related to the dispersion of their respective probability distributions. In fact, for $\alpha = 2$, the alpha-stable distribution is equivalent to a Gaussian distribution with variance $\sigma^2 = 2\gamma$. Observing the distribution of the parameter $\alpha$~in Figure~\ref{fig:alpha_trafego}, the similarity in the shape of the distributions of $\gamma$~ and $\sigma$ is justified, since most of the measured signals have an $\alpha$ value close to 2.

\begin{figure}[H] 
    \includegraphics[width=.95\linewidth]{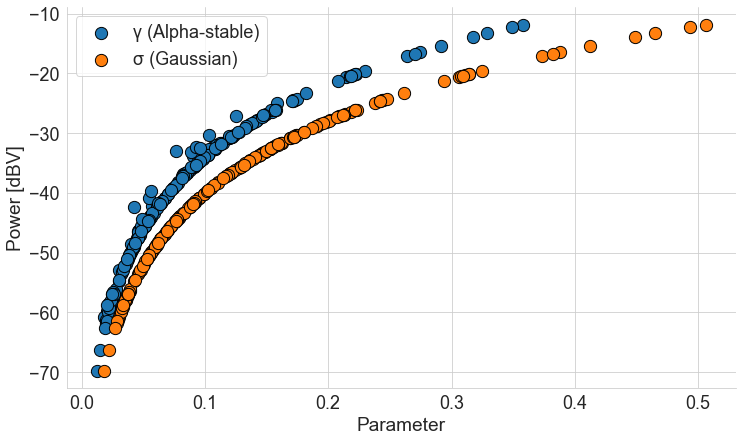}
    \caption{Relation between average power and $\gamma$~and $\sigma$. Curves were obtained using data from the first measurement campaign.}
    \label{fig:sigma_gamma_power}
\end{figure}

The relation between average power, $\gamma$, and $\sigma$ is illustrated in Figure~\ref{fig:sigma_gamma_power}, which graphs power in relation to the other parameters. To obtain the curves, only data from the first campaign were used, since a similar result can be shown for the second campaign. Both curves follow a logarithm shape, with oscillations in the case of $\gamma$. In the AWGN model, the noise variance is an estimator for its power. In contrast, variance and power are not defined for the alpha-stable distribution (unless $\alpha = 2$), meaning there is no direct association between the dispersion parameter and power such as in the Gaussian case. However, $\gamma$ is still helpful to measure the noise level in a stable noise model. For instance, some authors define a generalized version of the signal-to-noise ratio (GSNR) that takes into account the power of a $s(t)$ signal and the dispersion of an alpha-stable noise~\cite{Georgiou1999}: 
\begin{equation}
\label{eq:gsnr}
    GSNR = 10\log_{10} \left(\frac{1}{\gamma M} \sum_{t=1}^{M} |s(t)|^2 \right).
\end{equation}

Consequently, the results obtained for $\gamma$~and $\sigma$~reaffirm the conclusion that the fluidity of traffic and the speed at which the vehicle can move due to this traffic are related to the noise power levels received inside the vehicle.
\subsubsection{Window Analysis}

\begin{figure}[H] 
    \includegraphics[width=.95\linewidth]{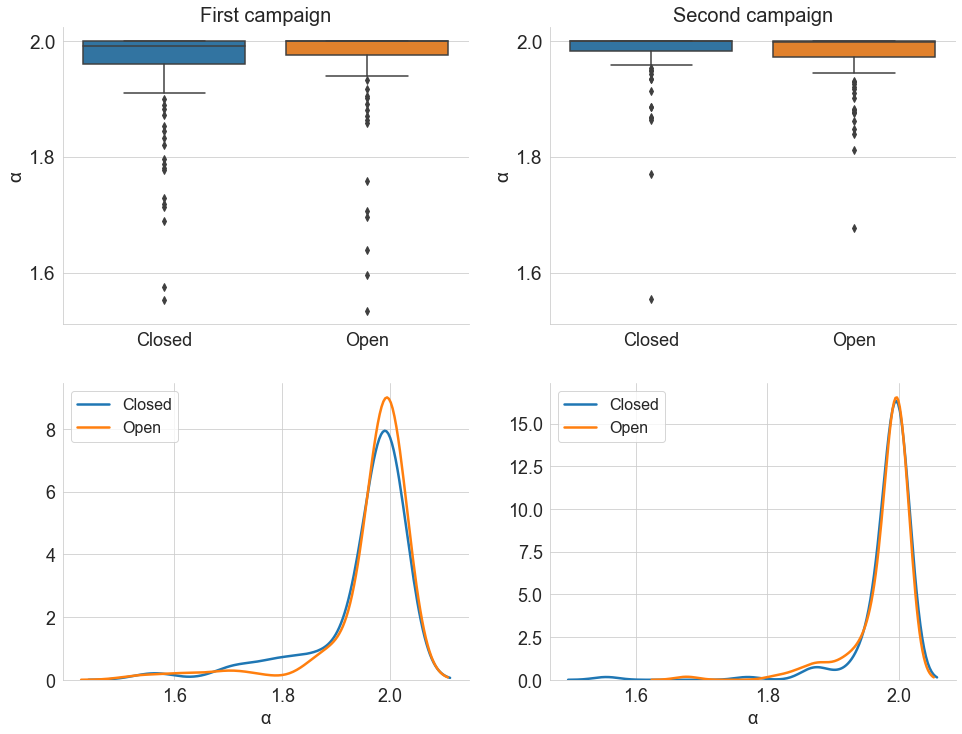}
    \caption{Distribution of the estimated parameter $\alpha$, from the alpha-stable distribution, grouped by \textit{Window}. The KDE curves were obtained using a Gaussian kernel.}
    \label{fig:alpha_window}
\end{figure}

\begin{figure}[H] 
    \includegraphics[width=.95\linewidth]{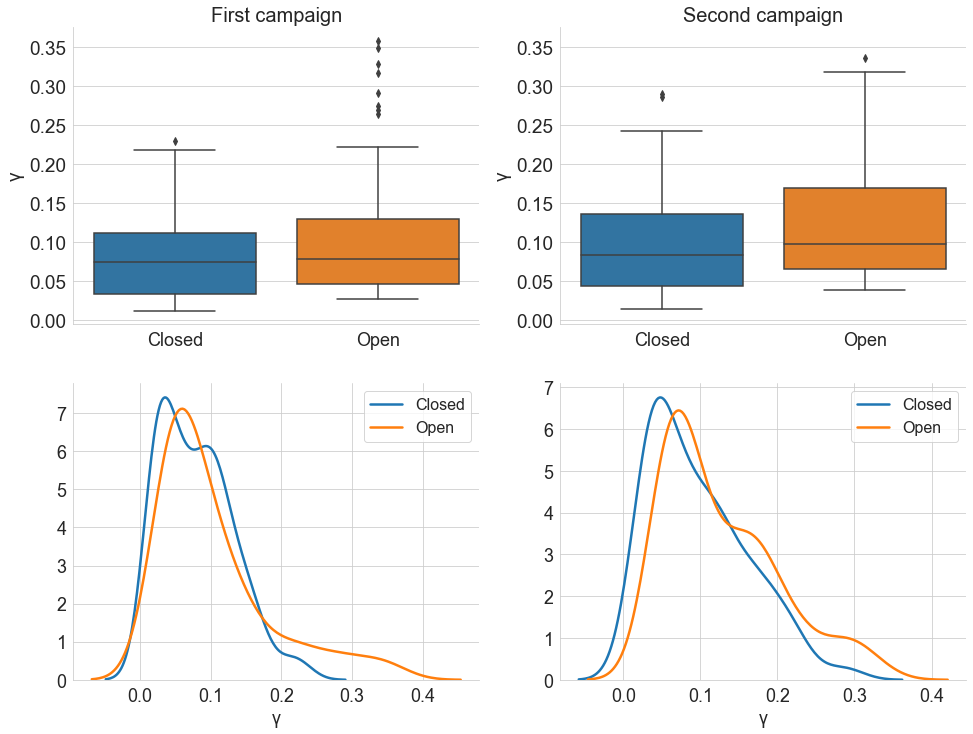}
    \caption{Distribution of the estimated parameter $\gamma$, from the alpha-stable distribution, grouped by \textit{Window}. The KDE curves were obtained using a Gaussian kernel.}
    \label{fig:gamma_window}
\end{figure}

\begin{figure}[H] 
    \includegraphics[width=.95\linewidth]{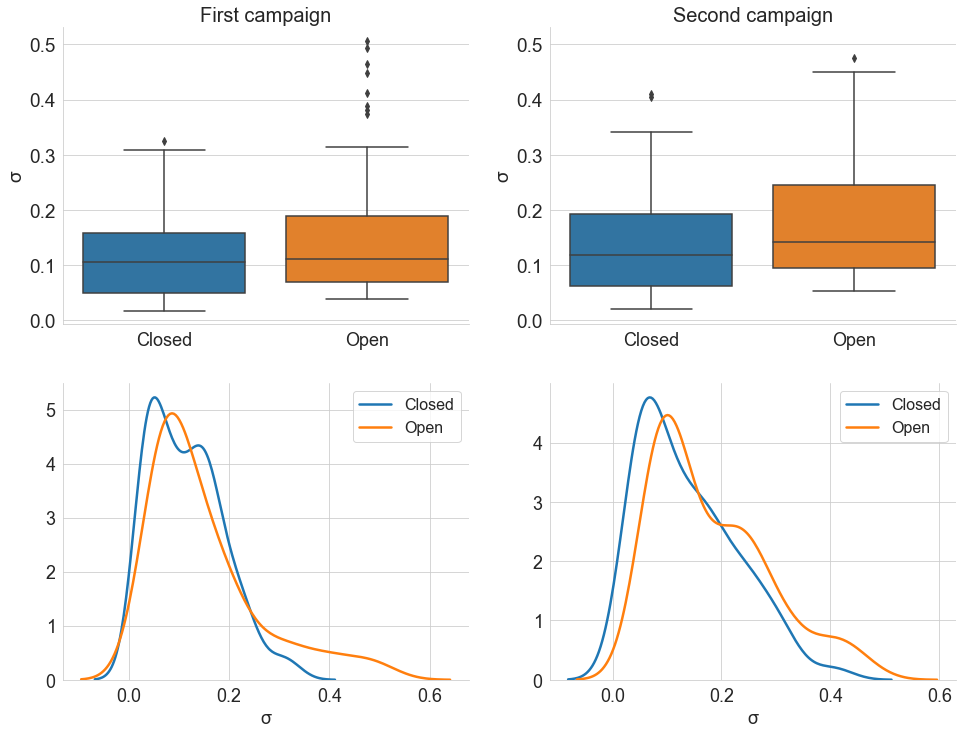}
    \caption{Distribution of the estimated parameter $\sigma$, from the alpha-stable distribution, grouped by \textit{Window}. The KDE curves were obtained using a Gaussian kernel.}
    \label{fig:sigma_window}
\end{figure}

Figures~\ref{fig:alpha_window}, \ref{fig:gamma_window} and \ref{fig:sigma_window} show, respectively, the distribution of the estimated parameters $\alpha$, $\gamma$, and $\sigma$, which are categorized by the \textit{Window} variable. In accordance with the previous section, most samples have an $\alpha$ value close to $2$ in both campaigns and window positions, and the dispersion is greater for the first campaign. However, unlike the \textit{Traffic} variable, the distribution of $\alpha$ is similar for both states of the windows. 

\vspace{-6pt}

Similarly, Figure~\ref{fig:gamma_window} indicates little difference between the distribution of $\gamma$ and $\sigma$ for the two positions of the car windows. There is only a trend of slightly higher values for the open windows scenario, which is a result compatible with the average power analysis. The results for the \textit{Window} and \textit{Traffic} variables lead to the conclusion that the position of the car's four windows has little influence on the level and degree of impulsiveness of the observed internal noise. From the standpoint of an audio reception system, external factors have less relevance to the internal noise composition when the vehicle is at higher speeds. It is important to emphasize that these conclusions are valid for the scenario described in Section~\ref{sec:campaign_conditions}.

\subsubsection{RMSE Evaluation}

Table~\ref{tab:rmse_fittings} lists the Root-Mean-Squared Error (RMSE) fitting errors of the models in all samples, comparing the performance of alpha-stable and Gaussian distributions. The first line shows that in both campaigns, more than half of the samples had a lower RMSE when modeled by an alpha-stable variable. In addition, the biggest absolute difference of RMSE when the Gaussian model performs better is 0.0060. In contrast, this difference for when the alpha-stable model has a superior performance is 1.1606 for Campaign 1 and 0.6124 for Campaign 2. Therefore, even when they are inferior in terms of RMSE, the alpha-stable models approximate the performance of Gaussian models, while the opposite does not~occur.

This is due to the greater flexibility of the alpha-stable variable. By adjusting the tail of its distribution, it is able to represent different degrees of impulsiveness, including the Gaussian case with $\alpha = 2$. Finally, the average RMSE of all samples is smaller for the alpha-stable option in both measurement groups. Hence, the alpha-stable distribution achieves a better fit in an overall assessment of the scenario.

\begin{table}[H] 
\caption{Comparison of RMSE values for the fitting of alpha-stable and Gaussian models..}
\label{tab:rmse_fittings}
\begin{tabular}{@{}ccc@{}}
\toprule
                                        & \textbf{First Campaign} & \textbf{Second Campaign} \\ \midrule
\begin{tabular}[c]{@{}c@{}}Proportion of samples with\\ a smaller RSME for the alpha-stable model\end{tabular}    & 71.13\% & 61.81\% \\ \midrule
\begin{tabular}[c]{@{}c@{}}Greatest difference in RSME\\ when alpha-stable model performs better\end{tabular}       & 1.1606   & 0.6124 \\ \midrule
\begin{tabular}[c]{@{}c@{}}Greatest difference in RSME\\ when Gaussian model performs better\end{tabular} & 0.0060   & 0.0060 \\ \midrule
Average RMSE for alpha-stable model      & 0.3136              & 0.1930              \\ \midrule
Average RMSE for Gaussian model& 0.3660              & 0.2064              \\ \bottomrule
\end{tabular}  

\end{table}

\subsection{Considerations about Window Size for Estimation}

The window size is an important parameter for any signal processing system. In the context of impulsive noise, the choice of the number of samples used to estimate the parameters of an alpha-stable probability distribution must take into account the trade-off between latency and stability of the estimation. Larger windows lead to faster convergence of parameters, while with smaller windows, the estimation tends to vary more depending on whether or not impulsive events are included in the observed interval. However, larger windows can represent a signaling cost that can make delay-sensitive applications~unfeasible.

\begin{figure}[H] 
    \includegraphics[width=0.95\linewidth]{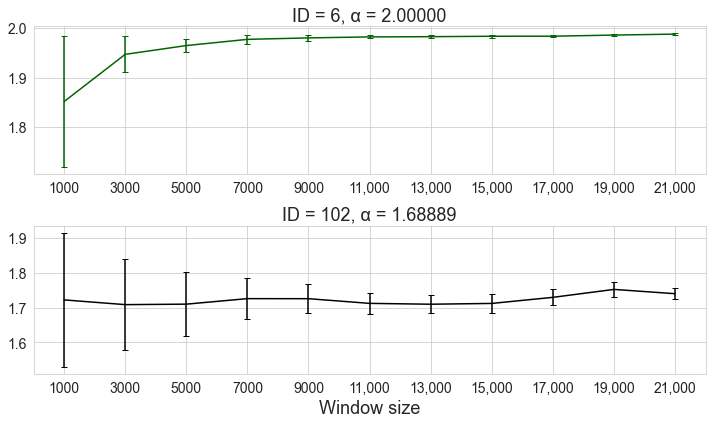}
    \caption{Estimation stability for different window sizes. The central value is the mean of the estimation, and the error bars are the variance. The first curve corresponds to a measurement from the \textit{Green} traffic group, and the second curve belongs to a \textit{Black} measurement.}
    \label{fig:window_variance}
\end{figure}

The effect of window size on the estimation of the $\alpha$ parameter by MLE can be seen in Figure~\ref{fig:window_variance}. 
Two of the measurements with 240,000 samples each were divided into fixed windows, ranging from 1000 to 21,000 samples with a step of 2000 samples. The figure shows the mean and variance of the estimated value of $\alpha$ for the computation of each window size. In the first curve, obtained from a measurement of the \textit{Green} traffic category, the estimated value rapidly converges to its final value, which is close to $\alpha = 2$, which is the value obtained when all samples are used. Likewise, the variation between the estimated values quickly becomes negligible. In the second curve, from a measurement of the \textit{Black} category, 
the variation in the estimated value is greater than for the first curve.

However, in both cases, the variance of the estimator decreases as the number of samples included in the window increases. This is a desirable feature for an estimator~\cite{scharf_1991}. The curves illustrate the trade-off between convergence to an optimal value and the processing time involved in the choice of window size. The convergence of the estimation depends on the degree of the impulsiveness of the measured signal and on the inclusion or not of impulsive events in the window~\cite{Danilo2019}. 

Another important factor to consider for windows with few samples is the possibility of numerical errors due to an insufficient amount of samples. For some of the noise samples tested for the convergence curves of $\alpha$, it was not possible to obtain results for a window size of 1000 samples. In these cases, the error is associated with the convergence of the parameter $\gamma$. As discussed in Section~\ref{sec:methods}, the use of MLE in MATLAB for the alpha-stable distribution depends on the choice of initial values to start the optimization routine. As these values are chosen based on quantiles of the signal, the number of samples in the window has a great influence on the result of the quantile estimation algorithm~\cite{Kristoffer2018}. In the case of the signals for which the error occurs, the optimizer is not able to find a valid value of $\gamma$ within the number of iterations imposed by the optimization routine, returning $\gamma = 0$. Therefore, the estimation for very small windows may be unfeasible. In the tests performed in this work, the smallest size for which all estimators converged to a valid result was for a window of 3000 samples.

\section{Final Remarks}\label{sec:conclusions}
\label{sec:conclusion}
Internal acoustic noise is an important factor in vehicle design. Interest in this topic grows due to concern with health and acoustic comfort issues and with the emergence of autonomous vehicles and new vehicular applications such as more advanced multimedia centers. Although there is extensive literature on the subject, most of the works focus on the study of the contribution to the noise of specific vehicle components in controlled environments such as test laboratories and under the perspective of psychoacoustics. In this work, an experimental evaluation of in-vehicle noise was presented, in which several noise samples were collected in two separate measurement campaigns in real traffic scenarios, with a setup elaborated from the perspective of multimedia systems and sound processing~applications.

The results found in both measurement campaigns show a strong correlation between the traffic level and the internal noise level. More fluid traffic, or equivalently a vehicle moving at higher speeds, results in higher average power levels. In contrast, the position of the car's windows showed a weak influence on the power level measured inside the vehicle. It is important to emphasize that this result, although counter-intuitive, was obtained from the perspective of an audio capture system. It was also shown that from the point of view of statistical modeling, speed and traffic are redundant variables, as the latter is defined from the former.

Moreover, a comparison of the AWGN and alpha-stable models was made in the modeling of the collected noise data. The comparison shows that models based on the stable distribution have a superior fit. An evaluation of the parameter $\alpha$ of the alpha-stable models revealed that although the internal noise has a predominant Gaussian characteristic, there is a relevant degree of impulsiveness. The frequency of these impulsive phenomena tends to be higher for heavier traffic situations. Thus, the alpha-stable model, particularly the $S\alpha S$ model, is an appropriate option for representing this type of noise~\cite{shao1993,Georgiou1999,Danilo2019}.

The results above lead to the conclusion that the most relevant factors for the characteristics of internal acoustic noise are internal to the vehicle. It is speculated that higher speeds lead to greater noise produced by the engine and other components of the car and by the wind, reducing the influence of other external factors, even when the vehicle's windows are open. In addition, the observed internal noise has significant impulsiveness, which again tends to have less relevance when the car is at higher speeds. Therefore, efforts to mitigate internal noise, such as studies related to active noise control and optimizing vehicle structure systems, should mind the noise produced by the vehicle itself, and the design of applications such as voice commands and source location should consider an impulsive noise model.

In future works, the authors consider evaluating the vehicular interior noise in the presence of speech signals, dealing with source separation systems.  In addition, we intend to extend our experiment with changes to the setup regarding the microphone's position and directivity. Such studies may provide new insights to comprehend and mitigate the noise in a vehicle interior environment, which will contribute to the development of in-vehicle voice and audio applications.

\vspace{6pt}
\authorcontributions{Conceptualization, D.F., D.P., V.A.d.S.J. and A.M.; Data curation, D.F., D.P., H.L.O. and L.P.; Formal analysis, D.F. and D.P.; Investigation, D.F., D.P., H.L.O. and L.P.; Methodology, D.F. and D.P.; Supervision, V.A.d.S.J. and A.M.; Visualization, D.F., H.L.O. and L.P.; Writing---original draft, D.F. and D.P.; Writing---review \& editing, V.A.d.S.J. and A.M. All authors have read and agreed to the published version of the manuscript.}

\funding{This study was financed in part by the Coordena\c{c}\~{a}o de Aperfei\c{c}oamento de Pessoal de N\'{i}vel Superior - Brasil (CAPES) - Finance Code 001.}

\dataavailability{The data collected and presented in this paper are available in~\cite{noise_data2019}.} 

\conflictsofinterest{The authors declare no conflict of interest.} 

\begin{adjustwidth}{-\extralength}{0cm} 

\reftitle{References}

\end{adjustwidth}

\begin{thebibliography}{999}

\bibitem[Kinsella(2019)]{wp_voicebot}
Kinsella, B.;~Mutchler, A.
\newblock \emph{In-Car Voice Assistant Consumer Report};
\newblock Technical report; Voicebot: Washington, DC, USA, 2019.


\bibitem[Che(2021)]{Chen2021}
Chen, S.; Gu, F.; Liang, C.; Meng, H. 
{Review on Active Noise Control Technology for $\alpha$-Stable Distribution
  Impulsive Noise}.
\newblock {\em Circuits Syst. Signal Process.} {\bf 2021}, \emph{41}, 1--38.
\newblock
  {\changeurlcolor{black}\href{https://doi.org/10.1007/s00034-021-01814-6}{\detokenize{https://doi.org/10.1007/s00034-021-01814-6}}}.

\bibitem[Flor \em{et~al.}(2020)Flor, Pena, Pena, A.~de Sousa, and
  Martins]{Flor_2020}
Flor, D.; Pena, D.; Pena, L.; A.~de Sousa, V.; Martins, A.
\newblock Characterization of Noise Level Inside a Vehicle under Different
  Conditions.
\newblock {\em Sensors} {\bf 2020}, {\em 20},~2471.
\newblock
  {\changeurlcolor{black}\href{https://doi.org/10.3390/s20092471}{\detokenize{https://doi.org/10.3390/s20092471}}}.

\bibitem[Pena \em{et~al.}(2019)Pena, Sousa, Pena, and
  de~Lucena~Flor]{noise_data2019}
Pena, D.; Sousa, V.; Pena, L.; de~Lucena~Flor, D.
\newblock Database of acoustic noise power levels in a vehicle interior under
  different conditions. Accessed in February, 2022. 
  
\newblock
  {\changeurlcolor{black}\href{https://doi.org/10.6084/m9.figshare.11370885.v1}{\detokenize{https://doi.org/10.6084/m9.figshare.11370885.v1}}}.

\bibitem[Basner \em{et~al.}(2014)Basner, Babisch, Davis, Brink, Clark, Janssen,
  and Stansfeld]{BASNER2014}
Basner, M.; Babisch, W.; Davis, A.; Brink, M.; Clark, C.; Janssen, S.;
  Stansfeld, S.
\newblock Auditory and non-auditory effects of noise on health.
\newblock {\em  Lancet} {\bf 2014}, {\em 383},~1325--1332.
\newblock
  {\changeurlcolor{black}\href{https://doi.org/10.1016/S0140-6736(13)61613-X}{\detokenize{https://doi.org/10.1016/S0140-6736(13)61613-X}}}.

\bibitem[Rudolph \em{et~al.}(2019)Rudolph, Shev, Paksarian, Merikangas,
  Mennitt, James, and Casey]{Kara2019}
Rudolph, K.E.; Shev, A.; Paksarian, D.; Merikangas, K.R.; Mennitt, D.J.; James,
  P.; Casey, J.A.
\newblock Environmental noise and sleep and mental health outcomes in a
  nationally representative sample of urban US adolescents.
\newblock {\em Environ. Epidemiol.} {\bf 2019}, {\em 3}, e056.

\bibitem[Sygna \em{et~al.}(2014)Sygna, Aasvang, Aamodt, Oftedal, and
  Krog]{SYGNA2014}
Sygna, K.; Aasvang, G.M.; Aamodt, G.; Oftedal, B.; Krog, N.H.
\newblock Road traffic noise, sleep and mental health.
\newblock {\em Environ. Res.} {\bf 2014}, {\em 131},~17--24.

\bibitem[Lercher \em{et~al.}(2003)Lercher, Evans, and Meis]{Gary2003}
Lercher, P.; Evans, G.W.; Meis, M.
\newblock Ambient Noise and Cognitive Processes among Primary Schoolchildren.
\newblock {\em Environ. Behav.} {\bf 2003}, {\em 35},~725--735.
\newblock
  {\changeurlcolor{black}\href{https://doi.org/10.1177/0013916503256260}{\detokenize{https://doi.org/10.1177/0013916503256260}}}.

\bibitem[Roswall \em{et~al.}(2017)Roswall, Raaschou-Nielsen, Ketzel,
  Gammelmark, Overvad, Olsen, and Sørensen]{ROSWALL2017}
Roswall, N.; Raaschou-Nielsen, O.; Ketzel, M.; Gammelmark, A.; Overvad, K.;
  Olsen, A.; Sørensen, M.
\newblock Long-term residential road traffic noise and NO$_2$ exposure in relation
  to risk of incident myocardial infarction---A Danish cohort study.
\newblock {\em Environ. Res.} {\bf 2017}, {\em 156},~80--86.
\newblock
  {\changeurlcolor{black}\href{https://doi.org/10.1016/j.envres.2017.03.019}{\detokenize{https://doi.org/10.1016/j.envres.2017.03.019}}}.

\bibitem[Vienneau \em{et~al.}(2015)Vienneau, Schindler, Perez, Probst-Hensch,
  and Röösli]{VIENNEAU2015}
Vienneau, D.; Schindler, C.; Perez, L.; Probst-Hensch, N.; Röösli, M.
\newblock The relationship between transportation noise exposure and ischemic
  heart disease: A meta-analysis.
\newblock {\em Environ. Res.} {\bf 2015}, {\em 138},~372--380.
\newblock
  {\changeurlcolor{black}\href{https://doi.org/10.1016/j.envres.2015.02.023}{\detokenize{https://doi.org/10.1016/j.envres.2015.02.023}}}.

\bibitem[Roswall \em{et~al.}(2018)Roswall, Raaschou-Nielsen, Jensen,
  Tjønneland, and Sørensen]{ROSWALL2018}
Roswall, N.; Raaschou-Nielsen, O.; Jensen, S.S.; Tjønneland, A.; Sørensen, M.
\newblock Long-term exposure to residential railway and road traffic noise and
  risk for diabetes in a Danish cohort.
\newblock {\em Environ. Res.} {\bf 2018}, {\em 160},~292--297.
\newblock
  {\changeurlcolor{black}\href{https://doi.org/10.1016/j.envres.2017.10.008}{\detokenize{https://doi.org/10.1016/j.envres.2017.10.008}}}.

\bibitem[Huang \em{et~al.}(2020)Huang, Wu, Huang, Yang, and Ding]{Huang2020_2}
Huang, H.; Wu, J.; Huang, X.; Yang, M.; Ding, W.
\newblock {A generalized inverse cascade method to identify and optimize
  vehicle interior noise sources}.
\newblock {\em J. Sound Vib.} {\bf 2020}, {\em 467},~115062.
\newblock
  {\changeurlcolor{black}\href{https://doi.org/10.1016/j.jsv.2019.115062}{\detokenize{https://doi.org/10.1016/j.jsv.2019.115062}}}.

\bibitem[Liu(2021)]{Liu2021}
{Numerical and experimental-based framework for vibro-acoustic coupling
  investigation on a vehicle door in the slamming event}.
\newblock {\em Mech. Syst. Signal Process.} {\bf 2021}, {\em
  158},~107759.
\newblock
  {\changeurlcolor{black}\href{https://doi.org/10.1016/j.ymssp.2021.107759}{\detokenize{https://doi.org/10.1016/j.ymssp.2021.107759}}}.

\bibitem[Horne \em{et~al.}(2021)Horne, Jashami, Monsere, Kothuri, and
  Hurwitz]{Horne2021}
Horne, D.; Jashami, H.; Monsere, C.M.; Kothuri, S.; Hurwitz, D.S.
\newblock {Evaluating In-Vehicle Sound and Vibration during Incursions on
  Sinusoidal Rumble Strips}.
\newblock {\em Transp. Res. Rec. J. Transp. Res. Board} {\bf 2021}, {\em 2675},~154--166.
\newblock
  {\changeurlcolor{black}\href{https://doi.org/10.1177/0361198120967947}{\detokenize{https://doi.org/10.1177/0361198120967947}}}.

\bibitem[Liu \em{et~al.}(2021)Liu, Sun, Wang, Sun, Li, and Guo]{Ningning2020}
Liu, N.; Sun, Y.; Wang, Y.; Sun, P.; Li, W.; Guo, H.
\newblock Mechanism of interior noise generation in high-speed vehicle based on
  anti-noise operational transfer path analysis.
\newblock {\em J. Automob. Eng.} {\bf 2021}, {\em 235},~273--287. 
  {\changeurlcolor{black}\href{https://doi.org/10.1177/0954407020937219}{\detokenize{https://doi.org/10.1177/0954407020937219}}}.

\bibitem[Lu \em{et~al.}(2021)Lu, Jen, and de~Klerk]{Lu2021}
Lu, M.H.; Jen, M.U.; de~Klerk, D.
\newblock {Case study: Separating source contributions of vehicle interior
  noise by operational transfer path analysis}.
\newblock {\em Noise Control. Eng. J.} {\bf 2021}, {\em 69},~39--52.
\newblock
  {\changeurlcolor{black}\href{https://doi.org/10.3397/1/37694}{\detokenize{https://doi.org/10.3397/1/37694}}}.

\bibitem[Deng \em{et~al.}(2020)Deng, Sun, and Li]{Jianghua2018}
Deng, J.; Sun, J.; Li, A.
\newblock Analysis and Optimization of Vehicle Interior Noise Caused by Tire
  Excitation.
\newblock  In \emph{Proceedings of China SAE Congress 2018: Selected Papers};
  (China~SAE); Springer: Singapore,  2020; pp.  723--735.

\bibitem[Almalkawi \em{et~al.}(2010)Almalkawi, {Guerrero Zapata}, Al-Karaki,
  and Morillo-Pozo]{Almalkawi2010}
Almalkawi, I.T.; {Guerrero Zapata}, M.; Al-Karaki, J.N.; Morillo-Pozo, J.
\newblock {Wireless Multimedia Sensor Networks: Current Trends and Future
  Directions}.
\newblock {\em Sensors} {\bf 2010}, {\em 10},~6662--6717.
\newblock
  {\changeurlcolor{black}\href{https://doi.org/10.3390/s100706662}{\detokenize{https://doi.org/10.3390/s100706662}}}.

\bibitem[Wang \em{et~al.}(2007)Wang, Ma, Ding, and Bi]{Wang2007}
Wang, X.; Ma, J.J.; Ding, L.; Bi, D.W.
\newblock {Robust Forecasting for Energy Efficiency of Wireless Multimedia
  Sensor Networks}.
\newblock {\em Sensors} {\bf 2007}, {\em 7},~2779--2807.
\newblock
  {\changeurlcolor{black}\href{https://doi.org/10.3390/s7112779}{\detokenize{https://doi.org/10.3390/s7112779}}}.

\bibitem[Jennehag \em{et~al.}(2016)Jennehag, Forsstrom, and
  Fiordigigli]{Jennehag2016}
Jennehag, U.; Forsstrom, S.; Fiordigigli, F.
\newblock {Low Delay Video Streaming on the Internet of Things Using Raspberry
  Pi}.
\newblock {\em Electronics} {\bf 2016}, {\em 5},~60.
\newblock
  {\changeurlcolor{black}\href{https://doi.org/10.3390/electronics5030060}{\detokenize{https://doi.org/10.3390/electronics5030060}}}.

\bibitem[Toma \em{et~al.}(2019)Toma, Alexandru, Popa, and Zamfiroiu]{Toma2019}
Toma.; Alexandru.; Popa.; Zamfiroiu.
\newblock {IoT Solution for Smart Cities' Pollution Monitoring and the Security
  Challenges}.
\newblock {\em Sensors} {\bf 2019}, {\em 19},~3401.
\newblock
  {\changeurlcolor{black}\href{https://doi.org/10.3390/s19153401}{\detokenize{https://doi.org/10.3390/s19153401}}}.

\bibitem[Kordov(2019)]{Kordov2019}
Kordov, K.
\newblock {A Novel Audio Encryption Algorithm with Permutation-Substitution
  Architecture}.
\newblock {\em Electronics} {\bf 2019}, {\em 8},~530.
\newblock
  {\changeurlcolor{black}\href{https://doi.org/10.3390/electronics8050530}{\detokenize{https://doi.org/10.3390/electronics8050530}}}.

\bibitem[Liu \em{et~al.}(2019)Liu, Han, Fang, and Ma]{Liu2019}
Liu, L.; Han, Z.; Fang, L.; Ma, Z.
\newblock {Tell the Device Password: Smart Device Wi-Fi Connection Based on
  Audio Waves}.
\newblock {\em Sensors} {\bf 2019}, {\em 19},~618.
\newblock
  {\changeurlcolor{black}\href{https://doi.org/10.3390/s19030618}{\detokenize{https://doi.org/10.3390/s19030618}}}.

\bibitem[{Kuhn} \em{et~al.}(1999){Kuhn}, {Jameel}, {Stumpfle}, and
  {Haddadi}]{Kuhn99}
{Kuhn}, T.; {Jameel}, A.; {Stumpfle}, M.; {Haddadi}, A.
\newblock Hybrid in-car speech recognition for mobile multimedia applications.
\newblock In Proceedings of the  1999 IEEE 49th Vehicular Technology Conference (Cat. No.99CH36363),  Houston, TX, USA,  16--20 May 1999;  Volume~3.
\newblock
  {\changeurlcolor{black}\href{https://doi.org/10.1109/VETEC.1999.778396}{\detokenize{https://doi.org/10.1109/VETEC.1999.778396}}}.

\bibitem[{Higuchi} \em{et~al.}(2017){Higuchi}, {Shinohara}, {Nakamura},
  {Mitsuyoshi}, {Tokuno}, {Omiya}, {Hagiwara}, and {Takano}]{Higuchi2017}
{Higuchi}, M.; {Shinohara}, S.; {Nakamura}, M.; {Mitsuyoshi}, S.; {Tokuno}, S.;
  {Omiya}, Y.; {Hagiwara}, N.; {Takano}, T.
\newblock An effect of noise on mental health indicator using voice.
\newblock In Proceedings of the  2017 International Conference on Intelligent Informatics and
  Biomedical Sciences (ICIIBMS),  Okinawa, Japan,  24--26 November 2017.  
\newblock
  {\changeurlcolor{black}\href{https://doi.org/10.1109/ICIIBMS.2017.8279690}{\detokenize{https://doi.org/10.1109/ICIIBMS.2017.8279690}}}.

\bibitem[{Khan} \em{et~al.}(2017){Khan}, {Akmal}, {Ali}, and {Naeem}]{Khan2017}
{Khan}, S.; {Akmal}, H.; {Ali}, I.; {Naeem}, N.
\newblock Efficient and unique learning of in-car voice control for engineering
  education.
\newblock In Proceedings of the  2017 International Multi-topic Conference (INMIC), Lahore, Pakistan,  24--26 November 2017; pp. 1--6.
\newblock
  {\changeurlcolor{black}\href{https://doi.org/10.1109/INMIC.2017.8289467}{\detokenize{https://doi.org/10.1109/INMIC.2017.8289467}}}.

\bibitem[Tremoulet \em{et~al.}(2019)Tremoulet, Seacrist, {Ward McIntosh}, Loeb,
  DiPietro, and Tushak]{Tremoulet2019}
Tremoulet, P.D.; Seacrist, T.; {Ward McIntosh}, C.; Loeb, H.; DiPietro, A.;
  Tushak, S.
\newblock {Transporting Children in Autonomous Vehicles: An Exploratory Study}.
\newblock {\em Hum. Factors J. Hum. Factors Ergon. Soc.} \bf{ 2019}, \it{62}, 278-287.
\newblock
  {\changeurlcolor{black}\href{https://doi.org/10.1177/0018720819853993}{\detokenize{https://doi.org/10.1177/0018720819853993}}}.

\bibitem[Chen(2019)]{Chen2019}
Chen, S.C.
\newblock {Multimedia for Autonomous Driving}.
\newblock {\em IEEE MultiMedia} {\bf 2019}, {\em 26},~5--8.
\newblock
  {\changeurlcolor{black}\href{https://doi.org/10.1109/MMUL.2019.2935397}{\detokenize{https://doi.org/10.1109/MMUL.2019.2935397}}}.

\bibitem[Doleschal \em{et~al.}(2021)Doleschal, Rottengruber, and
  Verhey]{DOLESCHAL2021}
Doleschal, F.; Rottengruber, H.; Verhey, J.L.
\newblock Influence parameters on the perceived magnitude of tonal content of
  electric vehicle interior sounds.
\newblock {\em Appl. Acoust.} {\bf 2021}, {\em 181},~108155.
\newblock
  {\changeurlcolor{black}\href{https://doi.org/10.1016/j.apacoust.2021.108155}{\detokenize{https://doi.org/10.1016/j.apacoust.2021.108155}}}.

\bibitem[Liao and Zheng(2020)]{LIAO2020}
Liao, X.; Zheng, S.
\newblock Quantification and characterization of the role of subjective
  preferences on vehicle acceleration sound quality.
\newblock {\em Mech. Syst. Signal Process.} {\bf 2020}, {\em
  138},~106549.
\newblock
  {\changeurlcolor{black}\href{https://doi.org/10.1016/j.ymssp.2019.106549}{\detokenize{https://doi.org/10.1016/j.ymssp.2019.106549}}}.

\bibitem[Swart and Bekker(2019)]{SWART2019}
Swart, D.; Bekker, A.
\newblock The relationship between consumer satisfaction and psychoacoustics of
  electric vehicle signature sound.
\newblock {\em Appl. Acoust.} {\bf 2019}, {\em 145},~167--175.
\newblock
  {\changeurlcolor{black}\href{https://doi.org/10.1016/j.apacoust.2018.09.019}{\detokenize{https://doi.org/10.1016/j.apacoust.2018.09.019}}}.

\bibitem[Huang \em{et~al.}(2018)Huang, Huang, Yang, Lim, and Ding]{Huang2018}
Huang, H.B.; Huang, X.R.; Yang, M.L.; Lim, T.C.; Ding, W.P.
\newblock {Identification of vehicle interior noise sources based on wavelet
  transform and partial coherence analysis}.
\newblock {\em Mech. Syst. Signal Process.} {\bf 2018}, {\em
  109},~247--267.
\newblock
  {\changeurlcolor{black}\href{https://doi.org/10.1016/j.ymssp.2018.02.045}{\detokenize{https://doi.org/10.1016/j.ymssp.2018.02.045}}}.

\bibitem[Jung \em{et~al.}(2019)Jung, Elliott, and Cheer]{Jung2019}
Jung, W.; Elliott, S.J.; Cheer, J.
\newblock {Local active control of road noise inside a vehicle}.
\newblock {\em Mech. Syst. Signal Process.} {\bf 2019}, {\em
  121},~144--157.
\newblock
  {\changeurlcolor{black}\href{https://doi.org/10.1016/j.ymssp.2018.11.003}{\detokenize{https://doi.org/10.1016/j.ymssp.2018.11.003}}}.

\bibitem[Vanherpe \em{et~al.}(2011)Vanherpe, Baresch, Lafon, and
  Bordji]{vanherpe2011}
Vanherpe, F.; Baresch, D.; Lafon, P.; Bordji, M.
\newblock {Wavenumber-Frequency Analysis of the Wall Pressure Fluctuations in
  the Wake of a Car Side Mirror.} In Proceedings of the 17th AIAA/CEAS Aeroacoustics Conference (32nd AIAA Aeroacoustics Conference), Portland, Oregon, 5--8 June 2011 . 
\newblock
  {\changeurlcolor{black}\href{https://doi.org/10.2514/6.2011-2936}{\detokenize{https://doi.org/10.2514/6.2011-2936}}}.

\bibitem[Praticò(2014)]{PRATICO2014}
Praticò, F.G.
\newblock On the dependence of acoustic performance on pavement
  characteristics.
\newblock {\em Transp. Res. Part D  Transp. Environ.} {\bf
  2014}, {\em 29},~79--87.
\newblock
  {\changeurlcolor{black}\href{https://doi.org/10.1016/j.trd.2014.04.004}{\detokenize{https://doi.org/10.1016/j.trd.2014.04.004}}}.

\bibitem[Pizzo \em{et~al.}(2020)Pizzo, Teti, Moro, Bianco, Fredianelli, and
  Licitra]{DELPIZZO2020}
Pizzo, A.D.; Teti, L.; Moro, A.; Bianco, F.; Fredianelli, L.; Licitra, G.
\newblock Influence of texture on tyre road noise spectra in rubberized
  pavements.
\newblock {\em Appl. Acoust.} {\bf 2020}, {\em 159},~107080.
\newblock
  {\changeurlcolor{black}\href{https://doi.org/10.1016/j.apacoust.2019.107080}{\detokenize{https://doi.org/10.1016/j.apacoust.2019.107080}}}.

\bibitem[Licitra \em{et~al.}(2017)Licitra, Teti, Cerchiai, and
  Bianco]{LICITRA2017}
Licitra, G.; Teti, L.; Cerchiai, M.; Bianco, F.
\newblock The influence of tyres on the use of the CPX method for evaluating
  the effectiveness of a noise mitigation action based on low-noise road
  surfaces.
\newblock {\em Transp. Res. Part D Transp. Environ.} {\bf
  2017}, {\em 55},~217--226.
\newblock
  {\changeurlcolor{black}\href{https://doi.org/10.1016/j.trd.2017.07.002}{\detokenize{https://doi.org/10.1016/j.trd.2017.07.002}}}.

\bibitem[Praticò and Anfosso-Lédée(2012)]{PRATICO2012}
Praticò, F.G.; Anfosso-Lédée, F.
\newblock Trends and Issues in Mitigating Traffic Noise through Quiet
  Pavements.
\newblock {\em Procedia-Soc. Behav. Sci.} {\bf 2012}, {\em
  53},~203--212.
\newblock
  {\changeurlcolor{black}\href{https://doi.org/10.1016/j.sbspro.2012.09.873}{\detokenize{https://doi.org/10.1016/j.sbspro.2012.09.873}}}.

\bibitem[Licitra \em{et~al.}(2015)Licitra, Cerchiai, Teti, Ascari, Bianco, and
  Chetoni]{LICITRA2015}
Licitra, G.; Cerchiai, M.; Teti, L.; Ascari, E.; Bianco, F.; Chetoni, M.
\newblock Performance Assessment of Low-Noise Road Surfaces in the Leopoldo
  Project: Comparison and Validation of Different Measurement Methods.
\newblock {\em Coatings} {\bf 2015}, {\em 5},~3--25.
\newblock
  {\changeurlcolor{black}\href{https://doi.org/10.3390/coatings5010003}{\detokenize{https://doi.org/10.3390/coatings5010003}}}.

\bibitem[Shin \em{et~al.}(2013)Shin, Park, and Lee]{Shin2013}
Shin, T.J.; Park, D.C.; Lee, S.K.
\newblock {Objective evaluation of door-closing sound quality based on
  physiological acoustics}.
\newblock {\em Int. J. Automot. Technol.} {\bf 2013}, {\em
  14},~133--141.
\newblock
  {\changeurlcolor{black}\href{https://doi.org/10.1007/s12239-013-0015-1}{\detokenize{https://doi.org/10.1007/s12239-013-0015-1}}}.

\bibitem[Parizet \em{et~al.}(2008)Parizet, Guyader, and Nosulenko]{Parizet2008}
Parizet, E.; Guyader, E.; Nosulenko, V.
\newblock {Analysis of car door closing sound quality}.
\newblock {\em Appl. Acoust.} {\bf 2008}, {\em 69},~12--22.
\newblock
  {\changeurlcolor{black}\href{https://doi.org/10.1016/j.apacoust.2006.09.004}{\detokenize{https://doi.org/10.1016/j.apacoust.2006.09.004}}}.

\bibitem[Li \em{et~al.}(2018)Li, Qiao, Yu, and Shi]{Li2018}
Li, Q.; Qiao, F.; Yu, L.; Shi, J.
\newblock {Modeling vehicle interior noise exposure dose on freeways:
  Considering weaving segment designs and engine operation}.
\newblock {\em J. Air  Waste Manag. Assoc.} {\bf
  2018}, {\em 68},~576--587.
\newblock
  {\changeurlcolor{black}\href{https://doi.org/10.1080/10962247.2017.1350213}{\detokenize{https://doi.org/10.1080/10962247.2017.1350213}}}.

\bibitem[Soeta and Shimokura(2017)]{Soeta2017}
Soeta, Y.; Shimokura, R.
\newblock {Sound quality evaluation of air-conditioner noise based on factors
  of the autocorrelation function}.
\newblock {\em Appl. Acoust.} {\bf 2017}, {\em 124},~11--19.
\newblock
  {\changeurlcolor{black}\href{https://doi.org/10.1016/j.apacoust.2017.03.015}{\detokenize{https://doi.org/10.1016/j.apacoust.2017.03.015}}}.

\bibitem[Morgan \em{et~al.}(2004)Morgan, Naganarayana, and
  Shankar]{christopher2004}
Morgan, C.D.; Naganarayana, B.P.; Shankar, S.
\newblock Comparative Evaluation of Seat Belt Retractor Websense Mechanism
  Rattle Noise.
\newblock {\em SAE Trans.} {\bf 2004}, {\em 113},~127--134.

\bibitem[YOSHIDA and INOUE(2017)]{YOSHIDA2017}
YOSHIDA, J.; INOUE, A.
\newblock {Road {\&} wind noise contribution separation using only interior
  noise having multiple sound sources---Accuracy improvement and permutation
  solution-}.
\newblock {\em Mech. Eng. J.} {\bf 2017}, {\em
  4},~1700165.
\newblock
  {\changeurlcolor{black}\href{https://doi.org/10.1299/mej.17-00165}{\detokenize{https://doi.org/10.1299/mej.17-00165}}}.

\bibitem[Talay and Altinisik(2019)]{Talay2019}
Talay, E.; Altinisik, A.
\newblock {The effect of door structural stiffness and flexural components to
  the interior wind noise at elevated vehicle speeds}.
\newblock {\em Appl. Acoust.} {\bf 2019}, {\em 148},~86--96.
\newblock
  {\changeurlcolor{black}\href{https://doi.org/10.1016/j.apacoust.2018.12.005}{\detokenize{https://doi.org/10.1016/j.apacoust.2018.12.005}}}.

\bibitem[Volandri \em{et~al.}(2018)Volandri, {Di Puccio}, Forte, and
  Mattei]{Volandri2018}
Volandri, G.; {Di Puccio}, F.; Forte, P.; Mattei, L.
\newblock {Psychoacoustic analysis of power windows sounds: Correlation between
  subjective and objective evaluations}.
\newblock {\em Appl. Acoust.} {\bf 2018}, {\em 134},~160--170.
\newblock
  {\changeurlcolor{black}\href{https://doi.org/10.1016/j.apacoust.2017.11.020}{\detokenize{https://doi.org/10.1016/j.apacoust.2017.11.020}}}.

\bibitem[Wang \em{et~al.}(2014)Wang, Shen, and Xing]{Wang2014}
Wang, Y.; Shen, G.; Xing, Y.
\newblock {A sound quality model for objective synthesis evaluation of vehicle
  interior noise based on artificial neural network}.
\newblock {\em Mech. Syst. Signal Process.} {\bf 2014}, {\em
  45},~255--266.
\newblock
  {\changeurlcolor{black}\href{https://doi.org/10.1016/j.ymssp.2013.11.001}{\detokenize{https://doi.org/10.1016/j.ymssp.2013.11.001}}}.

\bibitem[Pietila and Lim(2012)]{Pietila2012}
Pietila, G.; Lim, T.C.
\newblock {Intelligent systems approaches to product sound quality evaluations---A review}.
\newblock {\em Appl. Acoust.} {\bf 2012}, {\em 73},~987--1002.
\newblock
  {\changeurlcolor{black}\href{https://doi.org/10.1016/j.apacoust.2012.04.012}{\detokenize{https://doi.org/10.1016/j.apacoust.2012.04.012}}}.

\bibitem[Huang \em{et~al.}(2020)Huang, Huang, Wu, Yang, and Ding]{Huang2020}
Huang, X.; Huang, H.; Wu, J.; Yang, M.; Ding, W.
\newblock {Sound quality prediction and improving of vehicle interior noise
  based on deep convolutional neural networks}.
\newblock {\em Expert Syst. Appl.} {\bf 2020}, {\em 160},~113657.
\newblock
  {\changeurlcolor{black}\href{https://doi.org/10.1016/j.eswa.2020.113657}{\detokenize{https://doi.org/10.1016/j.eswa.2020.113657}}}.

\bibitem[Park and Kang(2020)]{Park_Kang_2020}
Park, J.H.; Kang, Y.J.
\newblock Evaluation Index for Sporty Engine Sound Reflecting Evaluators’
  Tastes, Developed Using K-means Cluster Analysis.
\newblock {\em Int. J. Automot. Technol.} {\bf 2020}, {\em
  21},~1379–1389.
\newblock
  {\changeurlcolor{black}\href{https://doi.org/10.1007/s12239-020-0130-8}{\detokenize{https://doi.org/10.1007/s12239-020-0130-8}}}.

\bibitem[Wang \em{et~al.}(2017)Wang, Guo, Feng, Ju, and Wang]{Wang2017}
Wang, Y.; Guo, H.; Feng, T.; Ju, J.; Wang, X.
\newblock {Acoustic behavior prediction for low-frequency sound quality based
  on finite element method and artificial neural network}.
\newblock {\em Appl. Acoust.} {\bf 2017}, {\em 122},~62--71.
\newblock
  {\changeurlcolor{black}\href{https://doi.org/10.1016/j.apacoust.2017.02.009}{\detokenize{https://doi.org/10.1016/j.apacoust.2017.02.009}}}.

\bibitem[{Puyana Romero} \em{et~al.}(2016){Puyana Romero}, Maffei, Brambilla,
  and Ciaburro]{PuyanaRomero2016}
{Puyana Romero}, V.; Maffei, L.; Brambilla, G.; Ciaburro, G.
\newblock {Acoustic, Visual and Spatial Indicators for the Description of the
  Soundscape of Waterfront Areas with and without Road Traffic Flow}.
\newblock {\em Int. J. Environ. Res. Public Health} {\bf 2016}, {\em 13},~934.
\newblock
  {\changeurlcolor{black}\href{https://doi.org/10.3390/ijerph13090934}{\detokenize{https://doi.org/10.3390/ijerph13090934}}}.

\bibitem[Bravo-Moncayo \em{et~al.}(2019)Bravo-Moncayo, Lucio-Naranjo,
  Ch{\'{a}}vez, Pav{\'{o}}n-Garc{\'{i}}a, and Garz{\'{o}}n]{Bravo-Moncayo2019}
Bravo-Moncayo, L.; Lucio-Naranjo, J.; Ch{\'{a}}vez, M.;
  Pav{\'{o}}n-Garc{\'{i}}a, I.; Garz{\'{o}}n, C.
\newblock {A machine learning approach for traffic-noise annoyance assessment}.
\newblock {\em Appl. Acoust.} {\bf 2019}, {\em 156},~262--270.
\newblock
  {\changeurlcolor{black}\href{https://doi.org/10.1016/j.apacoust.2019.07.010}{\detokenize{https://doi.org/10.1016/j.apacoust.2019.07.010}}}.

\bibitem[Yin \em{et~al.}(2020)Yin, Fallah-Shorshani, McConnell, Fruin, and
  Franklin]{Yin2020}
Yin, X.; Fallah-Shorshani, M.; McConnell, R.; Fruin, S.; Franklin, M.
\newblock {Predicting Fine Spatial Scale Traffic Noise Using Mobile
  Measurements and Machine Learning}.
\newblock {\em Environ. Sci. Technol.} {\bf 2020}, {\em
  54},~12860--12869.
\newblock
  {\changeurlcolor{black}\href{https://doi.org/10.1021/acs.est.0c01987}{\detokenize{https://doi.org/10.1021/acs.est.0c01987}}}.

\bibitem[Huang \em{et~al.}(2021)Huang, Wu, Lim, Yang, and Ding]{Huang2021}
Huang, H.; Wu, J.; Lim, T.C.; Yang, M.; Ding, W.
\newblock {Pure electric vehicle nonstationary interior sound quality
  prediction based on deep CNNs with an adaptable learning rate tree}.
\newblock {\em Mech. Syst. Signal Process.} {\bf 2021}, {\em
  148},~107170.
\newblock
  {\changeurlcolor{black}\href{https://doi.org/10.1016/j.ymssp.2020.107170}{\detokenize{https://doi.org/10.1016/j.ymssp.2020.107170}}}.

\bibitem[Qiu \em{et~al.}(2021)Qiu, Zhou, Xue, Tang, Wang, and Zhou]{Qiu2021}
Qiu, Y.; Zhou, E.; Xue, H.; Tang, Q.; Wang, G.; Zhou, B.
\newblock {Analysis on vehicle sound quality via deep belief network and
  optimization of exhaust system based on structure-SQE model}.
\newblock {\em Appl. Acoust.} {\bf 2021}, {\em 171},~107603.
\newblock
  {\changeurlcolor{black}\href{https://doi.org/10.1016/j.apacoust.2020.107603}{\detokenize{https://doi.org/10.1016/j.apacoust.2020.107603}}}.

\bibitem[Wang \em{et~al.}(2021)Wang, Li, Liu, Yang, Liu, and Xue]{Wang2021}
Wang, Z.; Li, P.; Liu, H.; Yang, J.; Liu, S.; Xue, L.
\newblock {Objective sound quality evaluation for the vehicle interior noise
  based on responses of the basilar membrane in the human ear}.
\newblock {\em Appl. Acoust.} {\bf 2021}, {\em 172},~107619.
\newblock
  {\changeurlcolor{black}\href{https://doi.org/10.1016/j.apacoust.2020.107619}{\detokenize{https://doi.org/10.1016/j.apacoust.2020.107619}}}.

\bibitem[Liu \em{et~al.}(2020)Liu, Lu, Zhang, Wang, Yin, and Hou]{Liu_2020}
Liu, Y.; Lu, W.; Zhang, Q.; Wang, X.; Yin, X.; Hou, H.
\newblock An Efficient Method for Prediction of the Flow-induced Vehicle
  Interior Noise.
\newblock {\em J. Phys. Conf. Ser.} {\bf 2020}, {\em
  1650},~022114.
\newblock
  {\changeurlcolor{black}\href{https://doi.org/10.1088/1742-6596/1650/2/022114}{\detokenize{https://doi.org/10.1088/1742-6596/1650/2/022114}}}.

\bibitem[Hägglund(2018)]{Kristoffer2018}
Hägglund, K.
\newblock {Symmetric alpha-Stable Adapted Demodulation and Parameter
  Estimation}.
\newblock Master's thesis, Lulea University of Technology,  Lulea, Sweden, 2018.

\bibitem[Nikias and Shao(1994)]{Nikias_1994}
Nikias, C.; Shao, M.
\newblock Recent Advances in Signal Processing with $\alpha$-Stable
  Distributions.  {\em IFAC Proc. Vol.} {\bf 1994}, {\em 27},~65--70. IFAC Symposium on System Identification (SYSID'94), Copenhagen,
  Denmark, 4-6 July. 
  {\changeurlcolor{black}\href{https://doi.org/10.1016/S1474-6670(17)47693-2}{\detokenize{https://doi.org/10.1016/S1474-6670(17)47693-2}}}.

\bibitem[Yardimci \em{et~al.}(1998)Yardimci, Cetin, and Cadzow]{Yardimici_1998}
Yardimci, Y.; Cetin, A.; Cadzow, J.
\newblock Robust direction-of-arrival estimation in non-Gaussian noise.
\newblock {\em IEEE Trans. Signal Process.} {\bf 1998}, {\em
  46},~1443--1451.
\newblock
  {\changeurlcolor{black}\href{https://doi.org/10.1109/78.668808}{\detokenize{https://doi.org/10.1109/78.668808}}}.

\bibitem[Georgiou(1999)]{Georgiou1999}
Georgiou, P.G.
\newblock {Alpha-stable modeling of noise and robust time-delay estimation in
  the presence of impulsive noise}.
\newblock {\em IEEE Trans. Multimed.} {\bf 1999}, {\em 1},~291--301.
\newblock
  {\changeurlcolor{black}\href{https://doi.org/10.1109/6046.784467}{\detokenize{https://doi.org/10.1109/6046.784467}}}.

\bibitem[Georgiou and Kyriakakis(2006)]{Georgiou_2006}
Georgiou, P.; Kyriakakis, C.
\newblock Robust maximum likelihood source localization: The case for
  sub-Gaussian versus Gaussian.
\newblock {\em IEEE Trans. Audio, Speech, Lang. Process.}
  {\bf 2006}, {\em 14},~1470--1480.
\newblock
  {\changeurlcolor{black}\href{https://doi.org/10.1109/TSA.2005.860846}{\detokenize{https://doi.org/10.1109/TSA.2005.860846}}}.

\bibitem[Guo \em{et~al.}(2021)Guo, Sun, Dai, and Chang]{Guo_2021}
Guo, M.; Sun, Y.; Dai, J.; Chang, C.
\newblock Robust DOA estimation for burst impulsive noise.
\newblock {\em Digit. Signal Process.} {\bf 2021}, {\em 114},~103059.
\newblock
  {\changeurlcolor{black}\href{https://doi.org/10.1016/j.dsp.2021.103059}{\detokenize{https://doi.org/10.1016/j.dsp.2021.103059}}}.

\bibitem[Figueredo \em{et~al.}(2020{\natexlab{a}})Figueredo, Pena, Filho,
  Dória, Martins, and Sousa~Jr]{Mario2020}
Figueredo, M.G.F.; Pena, D.d.S.; Filho, C.A.d.L.; Dória, M.F.d.S.; Martins,
  A.d.M.; Sousa~Jr, V.A.d.
\newblock Análise de Desempenho de Métodos de DOA Sujeitos a Modelos de Ruído Impulsivo com Misturas Gaussinas.  Os impactos de estudos voltados para as ciências exatas, Braz. Journals. \textbf{2020}, \textit{1}, 391-419. 

  
  
  
\newblock
  {\changeurlcolor{black}\href{https://doi.org/10.35587/brj.ed.0000606}{\detokenize{https://doi.org/10.35587/brj.ed.0000606}}}.

\bibitem[Figueredo \em{et~al.}(2020{\natexlab{b}})Figueredo, Pena, Lima~Filho,
  Dória, Martins, and Sousa~Jr]{Mario2020_2}
Figueredo, M.G.F.; Pena, D.d.S.; Lima~Filho, C.A.; Dória, M.F.d.S.; Martins,
  A.d.M.; Sousa~Jr, V.A.
\newblock  {Uma ferramenta de prototipagem para análise de técnicas de
  estimação de direção de chegada}.  \emph{Braz. J. Dev.} \textbf{2020}, \emph{6}, 28331--28342.  
\newblock
  {\changeurlcolor{black}\href{https://doi.org/10.34117/bjdv6n5-326}{\detokenize{https://doi.org/10.34117/bjdv6n5-326}}}.

\bibitem[Pena \em{et~al.}(2020)Pena, Lima, de~Sousa~Jr, Silveira, and
  Martins]{Danilo2020}
Pena, D.; Lima, A.; de~Sousa~Jr, V.; Silveira, L.; Martins, A.
\newblock Robust time delay estimation based on non-Gaussian impulsive acoustic
  channel.  \emph{J. Commun. Inf. Syst.} \textbf{2020}, \emph{35}, 86--89.
\newblock
  {\changeurlcolor{black}\href{https://doi.org/10.14209/jcis.2020.9}{\detokenize{https://doi.org/10.14209/jcis.2020.9}}}.

\bibitem[Liu \em{et~al.}(2017)Liu, Zhang, Zhou, Jing, and Truong]{Liu_2017}
Liu, H.; Zhang, R.; Zhou, Y.; Jing, X.; Truong, T.K.
\newblock Speech Denoising Using Transform Domains in the Presence of Impulsive
  and Gaussian Noises.
\newblock {\em IEEE Access} {\bf 2017}, {\em 5},~21193--21203.
\newblock
  {\changeurlcolor{black}\href{https://doi.org/10.1109/ACCESS.2017.2759142}{\detokenize{https://doi.org/10.1109/ACCESS.2017.2759142}}}.

\bibitem[Sasaoka \em{et~al.}(2020)Sasaoka, Akamatsu, Kawamara, Hayasaka, and
  ITOH]{Naoto_2020}
Sasaoka, N.; Akamatsu, E.; Kawamara, A.; Hayasaka, N.; ITOH, Y.
\newblock 4th Order Moment-Based Linear Prediction for Estimating Ringing Sound
  of Impulsive Noise in Speech Enhancement.
\newblock {\em IEICE Trans. Fundam. Electron. Commun. Comput. Sci.} {\bf 2020}, {\em E103.A},~1248--1251.
\newblock
  {\changeurlcolor{black}\href{https://doi.org/10.1587/transfun.2020EAL2005}{\detokenize{https://doi.org/10.1587/transfun.2020EAL2005}}}.

\bibitem[Rajala and Hongisto(2020)]{RAJALA_2020}
Rajala, V.; Hongisto, V.
\newblock Annoyance penalty of impulsive noise---The effect of impulse onset.
\newblock {\em Build. Environ.} {\bf 2020}, {\em 168},~106539.

\bibitem[Schaffeld \em{et~al.}(2020)Schaffeld, Schnitzler, Ruser, Woelfing,
  Baltzer, and Siebert]{Schaffeld_2020}
Schaffeld, T.; Schnitzler, J.G.; Ruser, A.; Woelfing, B.; Baltzer, J.; Siebert,
  U.
\newblock Effects of multiple exposures to pile driving noise on harbor
  porpoise hearing during simulated flights---An evaluation tool.
\newblock {\em  J. Acoust. Soc. Am.} {\bf 2020},
  {\em 147},~685--697.
\newblock
  {\changeurlcolor{black}\href{https://doi.org/10.1121/10.0000595}{\detokenize{https://doi.org/10.1121/10.0000595}}}.

\bibitem[Nolan(2020{\natexlab{a}})]{Nolan2020}
Nolan, J.P.
\newblock {\em Univariate Stable Distributions}; Springer:  Berlin/Heidelberg, Germany, 
  2020.
\newblock
  {\changeurlcolor{black}\href{https://doi.org/10.1007/978-3-030-52915-4}{\detokenize{https://doi.org/10.1007/978-3-030-52915-4}}}.

\bibitem[Nolan(2020{\natexlab{b}})]{Nolan_bib}
Nolan, J.P.
\newblock {Bibliography on stable distributions, processes and related topics}. 2006.


  

\bibitem[{Pokorný}(2017)]{Pokorny2017}
{Pokorný}, P.
\newblock Determining Traffic Levels in Cities Using Google Maps.
\newblock  In Proceedings of the 2017 Fourth International Conference on Mathematics and Computers in
  Sciences and in Industry (MCSI),  Corfu, Greece, 24--27 August 2017;  pp. 144--147.
\newblock
  {\changeurlcolor{black}\href{https://doi.org/10.1109/MCSI.2017.33}{\detokenize{https://doi.org/10.1109/MCSI.2017.33}}}.

\bibitem[Pena \em{et~al.}(2019)Pena, Lima, Dória, Pena, Martins, and
  Jr]{Danilo2019}
Pena, D.; Lima, C.; Dória, M.; Pena, L.; Martins, A.V., Jr.
\newblock Acoustic Impulsive Noise Based on Non-Gaussian Models: An
  Experimental Evaluation.
\newblock {\em Sensors} {\bf 2019}, {\em 19},~2827.
\newblock
  {\changeurlcolor{black}\href{https://doi.org/10.3390/s19122827}{\detokenize{https://doi.org/10.3390/s19122827}}}.

\bibitem[Respeaker(2019)]{respeaker_Online}
Respeaker.
\newblock {ReSpeaker 4 Mic Array for Raspberry Pi}.  2019.  Available online: 
\newblock \url{https://respeaker.io/4\_mic\_array/}  (accessed on 15 March 2021).

\bibitem[Stigler(1981)]{Stigler1981}
Stigler, S.M.
\newblock {Gauss and the Invention of Least Squares}.
\newblock {\em  Ann. Stat.} {\bf 1981}, {\em 9},~465--474.
\newblock
  {\changeurlcolor{black}\href{https://doi.org/10.1214/aos/1176345451}{\detokenize{https://doi.org/10.1214/aos/1176345451}}}.

\bibitem[Devore(2011)]{devore_2011}
Devore, J.L.
\newblock {\em Probability and Statistics for Engineering and the Sciences},
  8th ed.;  Cengage Learning: Boston, MA, USA,  2011.

\bibitem[Menard(2000)]{scott2000}
Menard, S.
\newblock Coefficients of Determination for Multiple Logistic Regression
  Analysis.
\newblock {\em  Am. Stat.} {\bf 2000}, {\em 54},~17--24.
\newblock
  {\changeurlcolor{black}\href{https://doi.org/10.1080/00031305.2000.10474502}{\detokenize{https://doi.org/10.1080/00031305.2000.10474502}}}.

\bibitem[Bai \em{et~al.}(2018)Bai, Tucci, Barmada, Raugi, and Zheng]{Bai2018}
Bai, L.; Tucci, M.; Barmada, S.; Raugi, M.; Zheng, T.
\newblock Impulsive Noise Characterization in Narrowband Power Line
  Communication.
\newblock {\em Energies} {\bf 2018}, {\em 11}.
\newblock
  {\changeurlcolor{black}\href{https://doi.org/10.3390/en11040863}{\detokenize{https://doi.org/10.3390/en11040863}}}.

\bibitem[Landa \em{et~al.}(2020)Landa, Díez-Bravo, Velez, and
  Arrinda]{Landa_2020}
Landa, I.; Díez-Bravo, A.; Velez, M.; Arrinda, A.
\newblock Influence of impulsive noise from lifts on orthogonal
  frequency-division multiplexing digital radio mondiale signals.
\newblock {\em IET Commun.} {\bf 2020}, {\em 14},~111--116.
\newblock
  {\changeurlcolor{black}\href{https://doi.org/10.1049/iet-com.2018.5458}{\detokenize{https://doi.org/10.1049/iet-com.2018.5458}}}.

\bibitem[Alam \em{et~al.}(2018)Alam, Kaddoum, and Agba]{Alam_2018}
Alam, M.S.; Kaddoum, G.; Agba, B.L.
\newblock Performance Analysis of Distributed Wireless Sensor Networks for
  Gaussian Source Estimation in the Presence of Impulsive Noise.
\newblock {\em IEEE Signal Process. Lett.} {\bf 2018}, {\em 25},~803--807.
\newblock
  {\changeurlcolor{black}\href{https://doi.org/10.1109/LSP.2018.2825951}{\detokenize{https://doi.org/10.1109/LSP.2018.2825951}}}.

\bibitem[Shao and Nikias(1993)]{shao1993}
Shao, M.; Nikias, C.
\newblock Signal processing with fractional lower order moments: stable
  processes and their applications.
\newblock {\em Proc. IEEE} {\bf 1993}, {\em 81},~986--1010.
\newblock
  {\changeurlcolor{black}\href{https://doi.org/10.1109/5.231338}{\detokenize{https://doi.org/10.1109/5.231338}}}.

\bibitem[Belkacemi and Marcos(2007)]{BELKACEMI_2007}
Belkacemi, H.; Marcos, S.
\newblock Robust subspace-based algorithms for joint angle/Doppler estimation
  in non-Gaussian clutter.
\newblock {\em Signal Process.} {\bf 2007}, {\em 87},~1547--1558.
\newblock
  {\changeurlcolor{black}\href{https://doi.org/10.1016/j.sigpro.2006.12.015}{\detokenize{https://doi.org/10.1016/j.sigpro.2006.12.015}}}.

\bibitem[Scharf(1991)]{scharf_1991}
Scharf, L.L.
\newblock {\em {Statistical Signal Processing-Detection, Estimation and Time
  Series Analysis}}, 5th ed.; Addison-Wesley Publishing Company: Boston, MA, USA, 1991.
\bibitem[Nolan(1997)]{Nolan1997}
Nolan, J.P.
\newblock Numerical calculation of stable densities and distribution functions.
\newblock {\em Commun. Stat. Stoch. Model.} {\bf 1997},
  {\em 13},~759--774.
\newblock
  {\changeurlcolor{black}\href{https://doi.org/10.1080/15326349708807450}{\detokenize{https://doi.org/10.1080/15326349708807450}}}.

\bibitem[Nolan(2001)]{Nolan2001}
Nolan, J.P., Maximum Likelihood Estimation and Diagnostics for Stable
  Distributions.
\newblock In {\em L{\'e}vy Processes: Theory and Applications};
  Barndorff-Nielsen, O.E.; Resnick, S.I.; Mikosch, T., Eds.; Birkh{\"a}user
  Boston: Boston, MA,  USA, 2001; pp. 379--400.

\bibitem[McCulloch(1986)]{McCulloch1986}
McCulloch, J.H.
\newblock Simple consistent estimators of stable distribution parameters.
\newblock {\em Commun. Stat.-Simul. Comput.} {\bf
  1986}, {\em 15},~1109--1136.
\newblock
  {\changeurlcolor{black}\href{https://doi.org/10.1080/03610918608812563}{\detokenize{https://doi.org/10.1080/03610918608812563}}}.

\end{thebibliography}
\end{document}